\documentclass[runningheads]{llncs}

\usepackage[T1]{fontenc}
\usepackage{graphicx}
\usepackage{amssymb}
\usepackage{amsfonts}
\usepackage{amsmath}
\usepackage{algorithm}
\usepackage{algorithmicx}
\usepackage{algpseudocode}

\renewcommand{\algorithmicrequire}{\textbf{Input:}}
\renewcommand{\algorithmicensure}{\textbf{Output:}}

\begin{document}
\title{Computing the minimal perimeter polygon \\
for digital objects in the triangular tiling \\
(Preprint, November 11, 2024)}
\author{Petra Wiederhold}

\institute{Centro de Investigaci\'on y de Estudios Avanzados (CINVESTAV-IPN) \\
Avenida I.P.N. 2508, Col. San Pedro Zacatenco, 07360 M\'exico, CDMX  \\
\email{petra.wiederhold@cinvestav.mx , pwiederhold@gmail.com} \\
}

\maketitle             

\begin{abstract}
This work presents an algorithm, together with its correctness proof, to determine the minimum perimeter polygon (MPP) for digital objects given as regular complexes in the triangular plane tiling. Such objects are edge-\-adja\-cen\-cy-\-connec\-ted sets of triangle tiles that have no end tiles, and the point set union of all their tiles forms a simple polygon. Nevertheless,  the boundary paths of the objects are not assumed to be simple. Then the MPP is a weakly simple polygon that coincides with the relative convex hull (i.e., geodesic hull) of a set $A$ with respect to a simple polygon $B$, where $A\subset B$, but $A$ is not necessarily a polygon, in fact it is generally not connected. Our MPP algorithm relies on constructing and iteratively constraining cones of visibility through forthcoming boundary tiles, it uses the structure of the canonical boun\-da\-ry path, the MPP frontier is the shortest polygonal curve following this path. We also propose a boundary tracing algorithm to obtain such paths from the objects.

\end{abstract}


\section{Introduction}
\label{sect:introduction}


As an alternative to pixels in the discrete plane $\mathbb{Z}^2$ or $(c\mathbb{Z})^2$ ($c\in\mathbb{R}$, $c>0$) which may be identified with a square tiling, the triangular and hexagonal grids and tilings have received attention for several decades for 2D digital image mo\-del\-ling and analysis. Triangular pixels have been used to study convexity properties in \cite{SklanskyKib1976}, in digital topology related to thinning in \cite{Saha2001,WiederholdMorales2009}, to develop thinning algorithms for adjacency graphs in \cite{Deutsch1972,Kardos2015,Kardos2017} and for cellular complexes in \cite{WiederholdMorales2009}, to consider topological curve problems in \cite{Nagy2024}, and, to design digital distance functions in \cite{Nagy2014} and geometric transformations in \cite{Nagy2018_Rotation,Nagy2020}. Coordinate systems to handle triangular pixels were proposed, for example, in \cite{Nagy2014}, and for triangular cell complexes in \cite{Nagy2015}. In \cite{Aman2023,Biswas2017}, digital objects in $\mathbb{Z}^2$ are approximated by polygons called triangular covers which are unions of tiles belonging to a triangular tiling superposed on $\mathbb{Z}^2$.
The present paper works with the plane tiling of equilateral triangles, which may serve as support of a digital image with triangular pixels, or, may represent more abstract discrete sets as those in \cite{Biswas2017}. 

In the context of studying convexity of digital plane objects, the \textit{mi\-ni\-mum pe\-ri\-me\-ter po\-ly\-gon (MPP)} was introduced in the 1970s in articles such as  \cite{Kim1982,SklanskyKib1976,SklanskyChazHans1972}, where pixels were identified with the tiles in polygonal tilings, certain assumptions guarantee that the MPP uniquely exists, and that all MPP vertices are vertices of the tiles. The latter is not true in other approaches where the MPP is determined as a result of an optimization problem \cite{Montanari1970}. In \cite{SklanskyKib1976,SklanskyChazHans1972}, two algorithms were proposed to determine the MPP vertices, based on the construction and iterative restriction of cones of visibility through forthcoming boundary tiles. The algorithm in \cite{SklanskyChazHans1972} is offered for sets of rectangular tiles. The more general algorithm in \cite{SklanskyKib1976} pretends to work for sets of tiles being convex polygons in so-called acute mosaics, where the union of any two edge-adjacent tiles forms a convex set; this is satisfied for rectangular and triangular tilings.

Much later, further studies on the MPP for the restricted class of 4-connected digital objects of square pixels (equivalently, for the square tiling) having simple contours, revealed that the MPP is useful to represent and to approximate digital objects, not only to describe convexity and concavity properties as discussed in \cite{Lachaud2011LinAlgorMLP,Roussillon2011}. As proved in \cite{SlobodaZatcoStoer1998} and reported in the textbook \cite{KletRosDigGeometry2004}, the MPP perimeter is a multigrid convergent perimeter estimator for any compact simply connected set $S\subset\mathbb{R}^2$ if its frontier is a smooth Jordan curve \cite{SlobodaStoer1994,SlobodaZatcoStoer1998}. In this regard, the MPP frontier is the minimal length Jordan curve which circumscribes the Inner Jordan digitization $J^-(S)$ (the union of all tiles lying inside $S$) but does not leave the Outer Jordan digitization $J^+(S)$ (the union of all tiles intersecting $S$) \cite{KletRosDigGeometry2004,SlobodaZatcoStoer1998}. The MPP in \cite{SlobodaZatcoStoer1998} was developed as the \textit{relative convex hull} of a set $A$ with respect to a set $B$ where both $A$ and $B$ are simple polygons and $A$ lies inside the interior of $B$. It was then applied to digitizations based on square pixels, the difference set $J^+(S)\setminus J^+(S)$ is a polyomino called \textit{simple grid-continuum}, it corresponds to a simple (that is, Jordan) 4-curve \cite{SlobodaStoer1994,SlobodaZatcoStoer1998}. For that special case, the MPP from \cite{SlobodaStoer1994,SlobodaZatcoStoer1998} coincides with the MPP due to Sklansky et al. \cite{SklanskyKib1976,SklanskyChazHans1972} which, at its origin, is more general and uses a distinct digitization. The relative convex hull also is considered in Computational Geometry and named \textit{geodesic hull} \cite{ToussaintGeodes}. 

For simple 4-curves, in particular, simple 4-contours, several efficient MPP algorithms were proposed which use strategies distinct from those applied in \cite{SklanskyKib1976,SklanskyChazHans1972}, for example, in \cite{Lachaud2011LinAlgorMLP,Roussillon2011,Provencal2009}. Some of these latter algorithms do not determine the MPP directly from the 4-contour, but from a collection of digital straight line segments of maximum length.

The MPP algorithms from \cite{SklanskyKib1976,SklanskyChazHans1972} are frequently cited in the literature, in particular, in the widely used modern textbooks \cite{GonzalezWoods4,GonzalezBookMatlab3} (2018, 2020), where adaptations to simple 4-contours are recommended to approximate objects in the digital plane $(c\mathbb{Z})^2$ \cite{KletteKovYip1999,KletRosDigGeometry2004}. 
The recent article \cite{Wiederhold2024} showed that both classical algorithms from \cite{SklanskyKib1976,SklanskyChazHans1972} and their adaptations and variants used, for example, in the books \cite{KletRosDigGeometry2004,GonzalezWoods4,GonzalezBookMatlab3}, all are failing, and that the mathematical foundation in \cite{SklanskyKib1976,SklanskyChazHans1972} contains serious errors, even for the restricted cases of square and rectangular tilings. We consider the general approach and many ideas of the algorithms from \cite{SklanskyKib1976,SklanskyChazHans1972} as interesting and worth to justify to restart investigations about the MPP in polygonal tilings, and to save the right original ideas. To this aim, a new MPP algorithm based on ideas of the classical works but correcting their errors, was presented in \cite{Wiederhold2024} for sets of rectangular tiles, where the correctness proof uses several properties that are specific for the rectangular tiling.

\begin{figure}
\centering
\includegraphics[height=2.6cm]{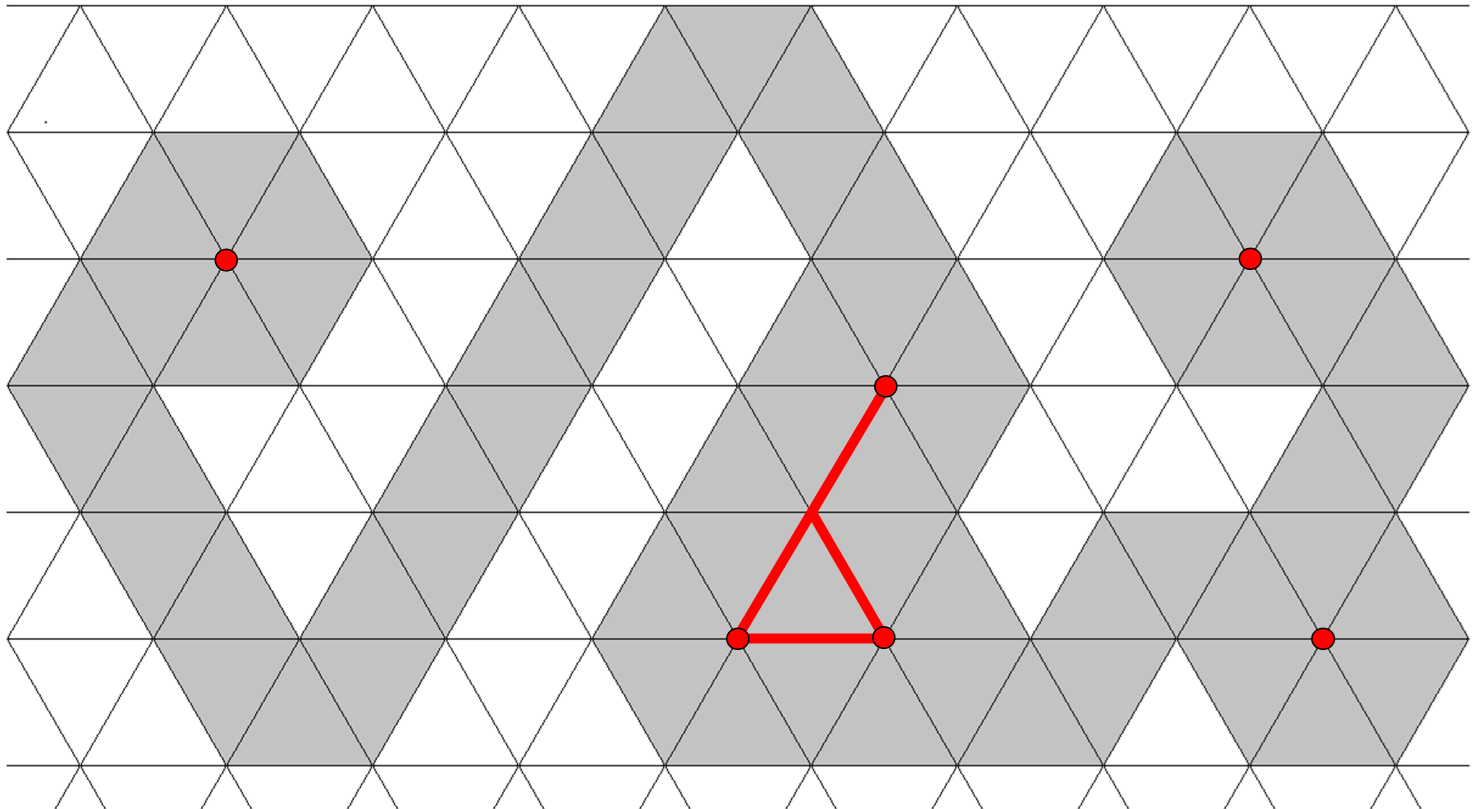}
\hskip0.5cm
\includegraphics[height=2.6cm]{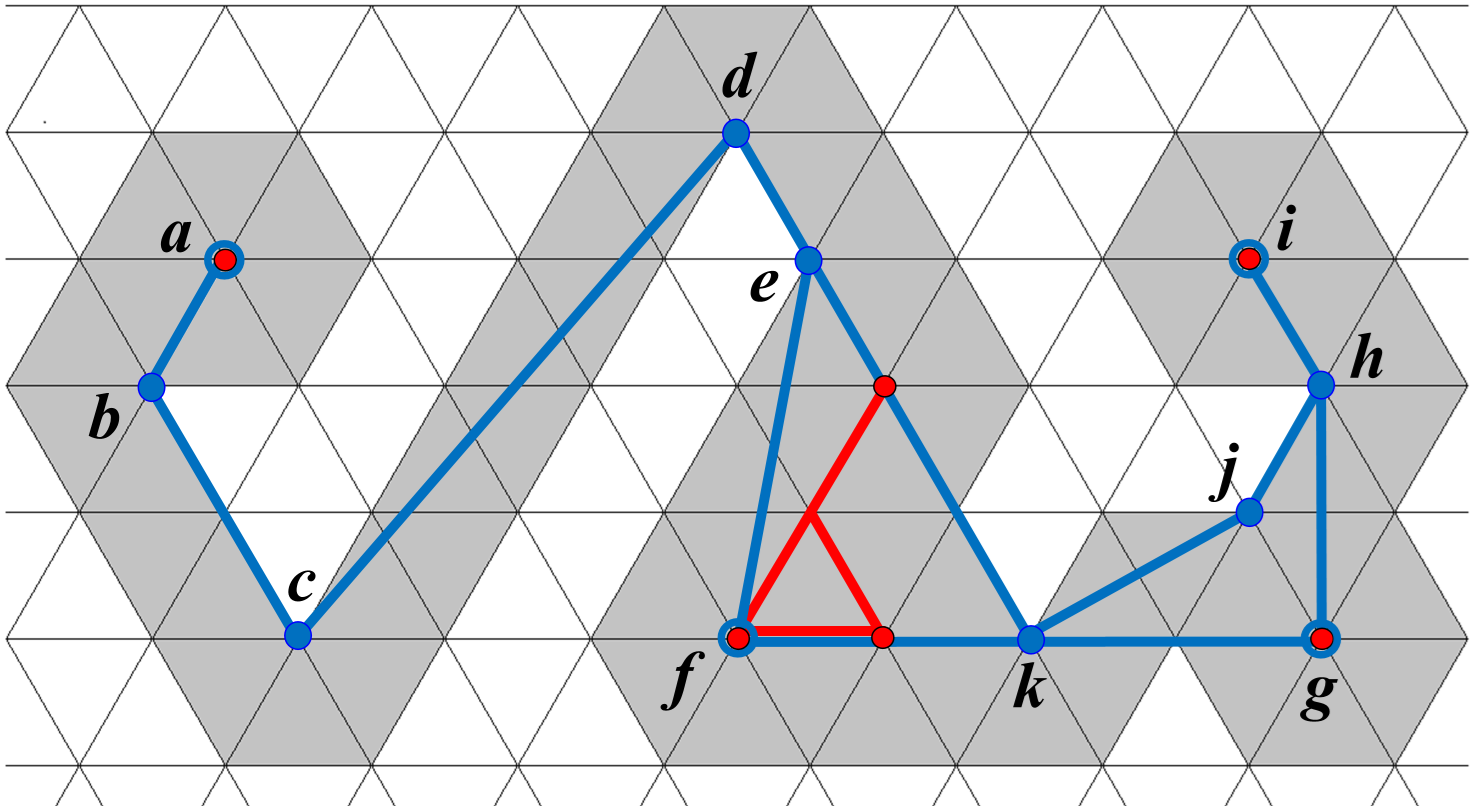}
\caption{A digital object $\mathcal{C}$ (shaded grey) given as a regular complex in the triangular tiling. The frontier of a special subset of $|\mathcal{C}|$ (the union of tiles belonging to $\mathcal{C}$), called the core of $\mathcal{C}$, is drawn in red. The right figure shows the MPP frontier depicted in blue, the MPP is the (weakly simple) polygon of shortest perimeter which contains the core and is contained in $|\mathcal{C}|$, it is given by the sequence $(a, b, c, d, e, f, g, h, i, h, j, k, d, c, b)$ of 15 vertices.}
\label{fig:Introd-example-regular-complex-MPP}
\end{figure}

The present article studies the MPP for digital objects made of triangular tiles given as regular complexes due to \cite{SklanskyKib1976,SklanskyChazHans1972} and proposes an algorithm to calculate the MPP vertices, together with illustrations by examples and a correctness proof. Such objects turn out to be edge-\-adja\-cency-\-connec\-ted sets of tiles that have no end tiles, but whose boundary paths are in general not simple. Figure \ref{fig:Introd-example-regular-complex-MPP} shows such an object with its MPP. The present paper generalizes and essentially extends the previous conference paper \cite{WiederholdLNCS2024}, which only considered the restricted case of objects having simple boundary paths.

Our MPP algorithm adapts the strategies from \cite{Wiederhold2024} to the triangular tiling, but the correctness proof requires new arguments due to the nature of triangular tiles. As input data we use the canonical boundary path obtained by boundary tracing, which is similar to known contour tracing for 4-connected objects made of square pixels. We also propose a boundary tracing algorithm for edge-\-adja\-cency-\-connec\-ted sets of triangle tiles to obtain such paths and show some properties. 

The MPP in this work finally coincides with the relative convex hull of a set $A$ with respect to a simple polygon $B$ which is a union of tiles. We have $A\subset B$, but $A$ may be far from being a simple polygon, in fact $A$ is not connected in general. It is important to note that our input data consist of a set of triangle tiles, for example triangular pixels. Although their union forms a polygon $B$, this polygon is not explicitly given as input. Moreover, although $B$ is a union of triangles, these triangles do not form a triangulation of the polygon $B$. Also, the information on the set $A$ is only implicitly included in the input data. Our MPP algorithm does not explicitly use the polygon $B$, in particular it is not based on a triangulation or other decomposition of $B$, but it takes advantage from the structure of the complex, specially of a boundary path which is easily determined as a cyclic sequence of triangle tiles. 

The article is organized as follows: Section \ref{sect:preliminaries} resumes preliminaries, Section \ref{sect:MPP-complexes} defines the MPP for sets of polygonal tiles due to \cite{SklanskyKib1976,SklanskyChazHans1972}. Section \ref{sect:boundary-path-from-tracing} presents a boun\-da\-ry tracing algorithm for the triangular tiling and studies properties of canonical boun\-da\-ry paths. Section \ref{sect:MPP-object-regular-complex} studies properties of regular objects in the triangular tiling. Section \ref{sect:MPP-Algorithm} proposes an MPP algorithm which is illustrated by examples. The correctness proof of the MPP algorithm is presented in Section \ref{sect:MPP-Alg-correctness}. Some Conclusions complete the paper.


\section{Preliminaries}
\label{sect:preliminaries}


We denote the set of integers by $\mathbb{Z}$, the Euclidean plane by $\mathbb{R}^2$, and the Euclidean distance by $d$. For $r\in \mathbb{R}$, $r>0$, and $p\in \mathbb{R}^2$, the set $U_r(p)= \{ q\in \mathbb{R}^2 : d(p,q) <r\}$ is an \textit{open disc} with centre $p$ and radius $r$. For $A\subset \mathbb{R}^2$, the (topological) \textit{frontier} of $A$ is the set $fr(A)$ of all points $p\in \mathbb{R}^2$ for which any $r\in \mathbb{R}$, $r>0$, satisfies that $U_r(p)$ intersects both $A$ and $(\mathbb{R}^2\setminus A)$.
For $p,q\in\mathbb{R}^2$, $\overline{pq}$ is the straight line segment from $p$ to $q$. If $p\neq q$, $\overrightarrow{pq}$ symbolizes the directed straight line segment from $p$ to $q$. All formulae and definitions are based on the standard Cartesian coordinate system in $\mathbb{R}^2$.

Recall that a curve is a set $\gamma =f([0,1])$ in $\mathbb{R}^2$, where $f:[0,1]\subset\mathbb{R} \rightarrow\mathbb{R}^2$ is a continuous function. The curve is closed if $f(0)= f(1)$. A closed curve $\gamma =f([0,1])$ is called a \textit{\textbf{simple curve}} or \textit{Jordan curve}\footnote{Intuitively, a Jordan curve consists of one piece, and, except that its starting point coincides with its end point, the curve does not touch itself.}
 if $f$ is injective on $[0,1)$.

In this paper, a \textbf{\textit{polygonal curve}} is understood as a closed curve $\gamma =f([0,1])\subset \mathbb{R}^2$, where there exist $n\in\mathbb{N}$ and $t_0$, $t_1$, $t_2,\cdots ,t_n$ with $t_0=0<t_1<\cdots <t_n=1$ such that for each $i\in\{ 1,2,\cdots ,n\}$, $f([t_{i-1},t_i])$ is a straight line segment. Then $\gamma$ has a finite \textit{length} which is the sum of lengths of all its straight line segments. A point $f(t_i)\in\gamma$ ($1\leq i\leq n-1$) is a \textit{\textbf{vertex}} of $\gamma$ if $f(t_i)\not\in \overline{f(t_{i-1})f(t_{i+1})}$). 
A \textbf{\textit{simple polygon}} is a compact connected subset of $\mathbb{R}^2$ whose frontier is a Jordan polygonal curve.

\begin{figure}
\centering
\includegraphics[height=2.1cm]{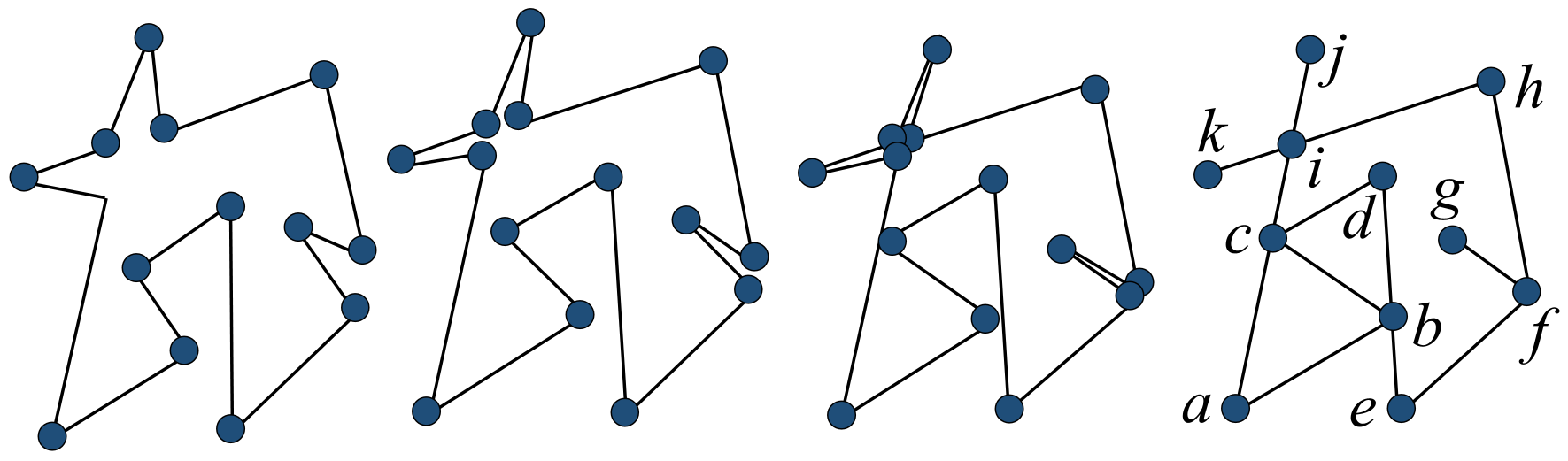}
\caption{From the left to the right: instances of a sequence of simple polygons converging to the weakly simple polygon with the cyclic sequence of vertices $(a,b,c,d,e,f,g,f,h,i,j,i,k,i)$.}
\label{fig:weak-simple-polyg}
\end{figure}

Due to \cite{SklanskyKib1976,SklanskyChazHans1972,Chang2014}, a polygonal curve $\gamma$ is called a \textbf{\textit{weakly simple polygonal curve}} if it is determined by an ordered sequence $(p_1,p_2, \cdots , p_k)$ of $k\geq 3$ vertices (that not necessarily are pairwise distinct), where for any $\epsilon >0$, there exists a simple po\-ly\-go\-nal curve $\delta $ determined by $k$ vertices $q_1,q_2, \cdots , q_k$ such that $d(p_j, q_j)<\epsilon $, being $d$ the Euclidean distance, for each $j\in\{ 1, 2, \cdots , k\}$. Moreover, any single point $p$ is defined to be a weakly simple polygonal curve having the unique vertex $p$, and, for distinct $p,q \in \mathbb{R}^2$, $\overline{pq}$ is a weakly simple polygonal curve given by the cyclic vertex sequence $(p,q)$. Then the polygon $P$ enclosed by $\gamma $ is called \textbf{\textit{weakly simple polygon}} and any vertex of $\gamma $ is a \textbf{\textit{vertex}} of $P$. 
A weakly simple polygon is enclosed by a curve $\gamma $ which may touch itself or trace back on itself but does not transversely cross itself \cite{ToussaintGeodes,Chang2014}. Such a curve $\gamma $ may be obtained as the limit of a sequence of Jordan curves $\gamma _j$ where for a finite number of vertices of $\gamma$, each such point is the limit of a sequence of corresponding vertices of the curves $\gamma _j$ \cite{SklanskyKib1976,SklanskyChazHans1972}, see Figure \ref{fig:weak-simple-polyg}.

By a \textbf{\textit{polygon}} we mean a compact subset of $\mathbb{R}^2$ which is a weakly simple polygon. In particular, any simple polygon is weakly simple. We assume that Jordan curves and weakly simple polygonal curves are always traced in counterclockwise sense. Then any polygon can be represented in a unique way as cyclic sequence of its vertices $(p_1, p_2, \cdots , p_k)$, where $p_i\not\in \overline{p_{(i-1\!\!\!\mod k)} p_{(i+1\!\!\!\mod k)}}$ for all $i\in\{ 1,\cdots , k\}$. The vertices of a simple polygon are pairwise distinct, which is not guaranteed for a weakly simple polygon, see Figure \ref{fig:weak-simple-polyg}.
The \textit{\textbf{perimeter}} of a polygon $P$ determined by the cyclic sequence $(p_1,p_2, \cdots , p_k)$ of vertices, $k\geq 1$, is given as the sum of the lengths of its frontier straight line segments: 
\[
\hbox{\textit{perimeter}}(P) = \sum_{i=1}^{i=k-1} d(p_i,p_{i+1})\  +\  d(p_k,p_1) \  .
\]
The polygon $P=\overline{pq}\, $ has perimeter $2 d(p,q)$, a polygon given as a single point has perimeter zero.

The orientation of a triple $p_1=(x_1,y_1)$, $p_2=(x_2,y_2)$, $p_3=(x_3,y_3)$ $\in \mathbb{R}^2$ can be described by the determinant
\[
D(p_1, p_2, p_3)=
\det \! \left( \!\!\! \begin{array}{ccc}
x_1 & y_1 & 1 \\
x_2 & y_2 & 1 \\
x_3 & y_3 & 1 \end{array} \!\!\! \right)  = x_1y_2+y_1x_3+x_2y_3-x_3y_2-x_2y_1-x_1y_3\, .
\]
In right-hand Cartesian coordinate systems, $D(p_1, p_2, p_3)<0$ if and only if $p_3$ lies on the right of $\overrightarrow{p_1p_2}$, i.e., $(p_1, p_2, p_3)$ forms a \textit{right turn}. $D(p_1, p_2, p_3)>0$ is equivalent to $p_3$ lying on the left of $\overrightarrow{p_1p_2}$, that is, $(p_1, p_2, p_3)$ forms a \textit{left turn}. $D(p_1, p_2, p_3)=0$ characterizes \textit{collinearity}, then $p_1,$ $p_2,$ $p_3$ belong to the same straight line segment which may be degenerated to a point. For consecutive points $p_1,$ $p_2,$ $p_3$ in a finite cyclic sequence of curve points of a polygonal curve $\gamma $ traced in counterclockwise sense, $p_2$ is named \textbf{\textit{convex}} if $D(p_1, p_2, p_3)>0$ (left turn), \textbf{\textit{concave}} if $D(p_1, p_2, p_3)<0$ (right turn), \textit{\textbf{linear point}} if $p_2\in\overline{p_1p_3}$, and a \textit{\textbf{peak}} if $D(p_1, p_2, p_3)=0$ but $p_2\not\in\overline{p_1p_3}$. While tracing a weakly simple polygonal curve $\gamma$ counterclockwise, the enclosed polygon $P$ lies on the left of (or on) $\gamma$, and $(\mathbb{R}^2\setminus P)$ lies strictly on its right side.


\section{Minimum perimeter polygon for complexes in polygonal tilings}
\label{sect:MPP-complexes}


A \textit{\textbf{polygonal tiling}} $\mathcal{M}$ is a family of convex polygons, called \textit{\textbf{tiles}}, each of which has a non-\-empty interior, whose union covers the plane, whose interiors are pairwise disjoint, and where for every $p\in\mathbb{R}^2$ there is an open disk with centre $p$ that meets only a finite number of tiles \cite{Gruenbaum,Schulte1993}. The intersection of two different tiles is either empty, or a non-\-zero straight line segment, called the \textit{\textbf{edge}} of the tiling, or a point that is a vertex of at least one of these tiles. A tiling is called \textit{\textbf{edge-to-edge}} \cite{Gruenbaum} if every edge belonging to different tiles $T_1, T_2$ coincides with a full side of $T_1$ and of $T_2$.
It is well-\-known that the \textit{square, triangular and hexagonal tiles} are the only possible edge-to-edge polygon tiles whose tiles are regular polygons of the same type. The \textit{rectangular tiling} is also edge-to-edge, its tiles are rectangles that do not necessarily have the same size.

In a polygonal tiling, two distinct tiles are called \textit{\textbf{(edge-) ad\-ja\-cent}} or \textbf{\textit{neigh\-bors}} if they share an edge. For the square tiling, this adjacency coincides with the 4-neighbor relation used in the discrete plane $(c\mathbb{Z})^2$, $c\in\mathbb{R}$, when each pixel $p\in (c\mathbb{Z})^2$ is identified with the square centered at $p$. Similarly as the well-known \textit{\textbf{4-neighborhood graph}} provides 4-paths and 4-connectivity \cite{KletRosDigGeometry2004,WiederholdEncyc2016}, a polygonal tiling $\mathcal{M}$ together with the adjacency relation yields the \textbf{\textit{adjacency graph}} $G(\mathcal{M})$ which provides paths and connec\-ti\-vi\-ty:  a \textit{\textbf{path}} (called \textit{chain} in \cite{SklanskyKib1976,SklanskyChazHans1972}) is a sequence of tiles $(t_1, t_2, \cdots , t_k)$ such that $t_i$ is a neighbor of $t_{i+1}$ for each $i=1,2, \cdots , n-1$, it is a \textit{\textbf{closed path}} if also $t_n$ is a neighbor of $t_1$. A subset $\mathcal{C}\subset\mathcal{M}$ is \textbf{\textit{(edge-\-ad\-ja\-cency-) connected}} if any two tiles of $\mathcal{C}$ are connected via a path whose elements all belong to $\mathcal{C}$. 

\begin{figure}
\centering
\includegraphics[height=3.7cm]{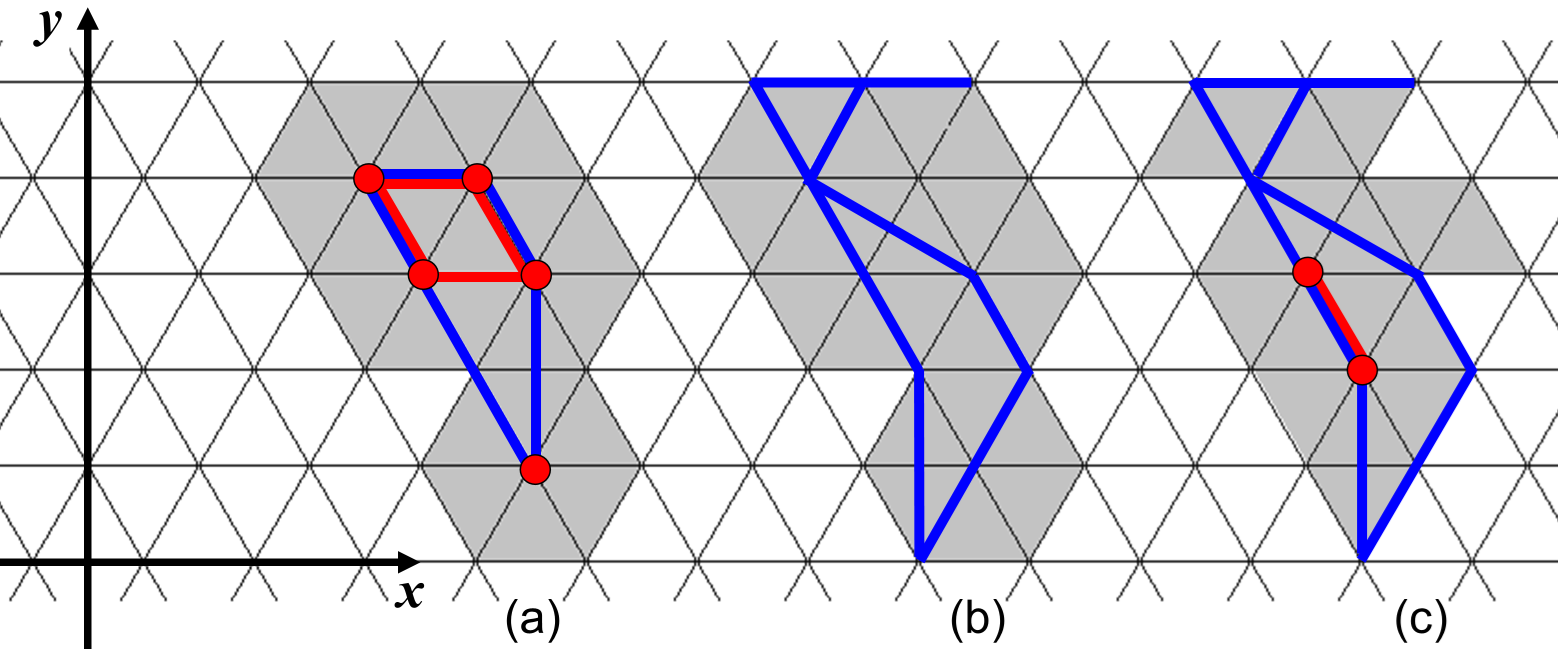}
\caption{A tiling of equilateral triangular tiles aligned to the Cartesian coordinate system. (a) and (b) present the same regular complex (shaded grey) whose core frontier is shown in red and the MPP frontier in blue in (a). This complex is also an image of the polygon drawn in blue in (b). The latter polygon is also a preimage of the complex shown in (c) which is neither regular nor normal (it has end tiles), its core is depicted in red.}
\label{fig:ImagePreimage}
\end{figure}

In the present work, a \textbf{\textit{triangular tiling}} is any edge-to-edge polygonal tiling $\mathcal{M}$ whose tiles are all equilateral triangles, then all tiles are of the same size. Since each tile has three neighbors, each node in the adjacency graph $G(\mathcal{M})$ has valence three. We suppose that the tiling is positioned as in Figure \ref{fig:ImagePreimage}, where the tiles form rows parallel to the $x$-axis of the Cartesian coordinate system. There are two types of tiles: upright and inverted triangles. Since all neighbors of an upright triangle are inverted triangles, and vice versa, the tiles in any path are alternately upright and inverted triangles.

Now let $\mathcal{M}$ be any polygonal tiling. We need to consider special types of paths in $\mathcal{M}$ and of subsets of $\mathcal{M}$ as follows:

\noindent $\bullet\  $ A path $(t_1, t_2, \cdots , t_k)$ is a \textit{\textbf{regular path}} if $t_{i-1}\neq t_{i+1}$ for each $2\leq i \leq k-1$ (and $t_{k-1}\neq t_1$, $t_k\neq t_2$, for a closed path) \cite{SklanskyKib1976,SklanskyChazHans1972}. A closed path is named \textit{\textbf{simple (closed) path}} if each of its tiles has exactly two neighbors in this path: this is a kind of \textit{Jordan digital curve} \cite{KletRosDigGeometry2004,WiederholdEncyc2016}. Every simple closed path is regular, but not vice versa. A path $(t_1, t_2, \cdots , t_k)$ with end tiles $t_1, t_k$ (that is, each of the tiles $t_1, t_k$ has exactly one neighbor in that path), is a \textit{\textbf{simple (open) path}} if each of its tiles, except $t_1$ and $t_k$, has exactly two neighbors in this path. 

\noindent $\bullet\  $ As in \cite{SklanskyKib1976,SklanskyChazHans1972}, let a \textit{\textbf{complex}} be any finite non-empty subset $\mathcal{C}$ of $\mathcal{M}$, denote its point set union by $|\mathcal{C}| = \bigcup \mathcal{C} = \{ p\in\mathbb{R}^2: p\in T\  \hbox{for a tile}\  T\in \mathcal{C}\} \subset \mathbb{R}^2$. If $\mathcal{C}$ has at least two elements, $T\in\mathcal{C}$ is an \textit{\textbf{end tile}} if $T$ is adjacent to exactly one other tile of $\mathcal{C}$. A \textbf{\textit{regular complex}} is a complex $\mathcal{C}$ that has a regular boun\-da\-ry path and where $fr(|\mathcal{C}|)$ is a Jordan curve. In the present work, any regular complex $\mathcal{C}$ in a triangular tiling is considered as a \textit{\textbf{digital object}}, then $\mathcal{C}$ is (edge-\-adja\-cen\-cy-) connected and has no end tile, and it is easy to see that $|\mathcal{C}|$ is a simple polygon.

\noindent $\bullet\  $ The \textbf{\textit{boun\-da\-ry}} $\mathcal{B}(\mathcal{C})$ of a complex $\mathcal{C}$ is defined as the set of tiles of $\mathcal{C}$ that meet the frontier $fr(|\mathcal{C} |)$. Any closed path that consists exactly of all tiles of $\mathcal{B}(\mathcal{C})$ is named a \textbf{\textit{boun\-da\-ry path}} of $\mathcal{C}$. While the set $\mathcal{B}(\mathcal{C})$ is uniquely defined, $\mathcal{C}$ may have several boun\-da\-ry paths with distinct properties, $\mathcal{B}(\mathcal{C})$ can even result as not connected, in which case $\mathcal{C}$ has no boun\-da\-ry path.

\noindent $\bullet\  $ The \textbf{\textit{core}} of a complex $\mathcal{C}$, denoted by $\hbox{\textit{core}}(\mathcal{C})\subset \mathbb{R}^2$, is the point set union of all tiles of $\mathcal{C}$, all edges (i.e., sides) belonging to tiles of $\mathcal{C}$, and all vertices of tiles of $\mathcal{C}$, provided that they do not meet the frontier $fr(|\mathcal{C}|)$ \cite{SklanskyKib1976}. 
Even for a connected complex $\mathcal{C}$, its core not necessarily is a polygon, it may even be disconnected, but its connected components are weakly simple polygons, see Figure \ref{fig:ImagePreimage}. All vertices of all connected components of $\hbox{\textit{core}}(\mathcal{C})$ are considered as \textbf{\textit{vertices of the core}}. Specially, the two endpoints of each component being a straight line segment and each isolated point of $\hbox{\textit{core}}(\mathcal{C})$ are considered as convex vertices of $\hbox{\textit{core}}(\mathcal{C})$.

\begin{definition} (from \cite{SklanskyKib1976,SklanskyChazHans1972})
Let $\mathcal{C}\subseteq\mathcal{M}$ be a complex in a polygonal tiling $\mathcal{M}$ and $P\subset\mathbb{R}^2$ a polygon.

\noindent \textbf{--} $C$ is called an \textbf{\textit{image}} of a $P$, and $P$ is called a \textbf{\textit{preimage}} of $\mathcal{C}$, if $P\subseteq |\mathcal{C}|$ and $T\cap P\neq\emptyset$ for each tile $T\in\mathcal{C}$.

\noindent \textbf{--} Any polygon that has shor\-test peri\-me\-ter among all poly\-gons which are pre\-images of $\, \mathcal{C}$ is called a \textbf{\textit{mini\-mum peri\-meter polygon (MPP)}} of $\, \mathcal{C}$.
\label{def:image-preimage-MPP}
\end{definition}

By Definition \ref{def:image-preimage-MPP}, an MPP is a polygon, so its frontier is a weakly simple polygonal curve, Figure \ref{fig:ImagePreimage} illustrates Definition \ref{def:image-preimage-MPP} for a triangular tiling. In general, the MPP of $\mathcal{C}$ is not unique, and its vertices are not guaranteed to be vertices of the tiles of $\mathcal{C}$. The existence and uniqueness of the MPP was proved for regular complexes in rectangular tilings \cite{SklanskyChazHans1972} and for normal complexes in  \textit{acute mosaics}, which are polygonal tilings where the union of each two adjacent tiles is convex \cite{SklanskyKib1976}. 
For this, a \textbf{\textit{normal complex}} is defined to have no end tile, and, where $|\mathcal{C}|$ is a simply connected set in $\mathbb{R}^2$ \cite{SklanskyKib1976}. A normal complex does not have to be edge-\-adja\-cency-\-connected, it can have so-called \textit{cut points} whose deletion disconnects $|\mathcal{C}|$. Clearly regular complexes are normal, but they have no cut points. 

As a consequence of the properties shown in \cite{SklanskyKib1976}, the MPP of any regular complex in a triangular tiling exists and is unique, see Section \ref{sect:MPP-object-regular-complex}. The MPP algorithm presented in Section \ref{sect:MPP-Algorithm} essentially uses a special type of boundary path of the complex. The next section studies properties of such paths and presents a method for determining them.


\section{Canonical boun\-da\-ry path from boun\-da\-ry tracing}
\label{sect:boundary-path-from-tracing}


\textit{Boun\-da\-ry tracing}, also called \textit{boun\-da\-ry following} or \textit{contour tracing}, is a standard method in digital image analysis to find the frontier of an object. It is described for 4- and 8-connected objects in $\mathbb{Z}^2$, for example, in the textbooks \cite{GonzalezWoods4,KletRosDigGeometry2004,GonzalezBookMatlab3} and in \cite{WiederholdVillafuerte2022}. Boun\-da\-ry tracing was generalized in \cite{Wiederhold2024} to objects given as edge-\-adja\-cen\-cy-\-connec\-ted complexes in rectangular tilings.

\begin{figure}
\centering
\includegraphics[height=4cm]{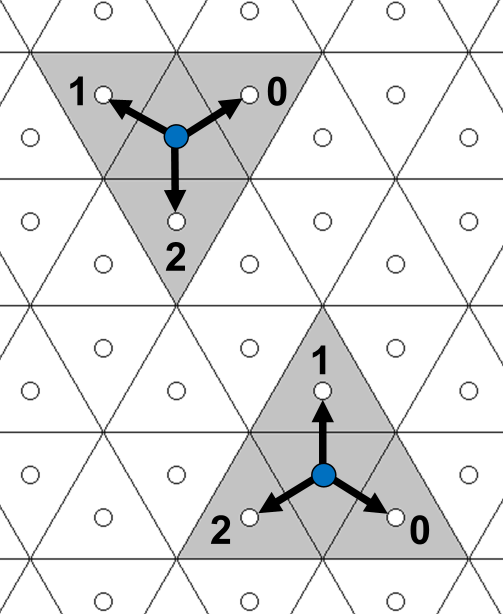}
\hskip0.1cm
\includegraphics[height=3.9cm]{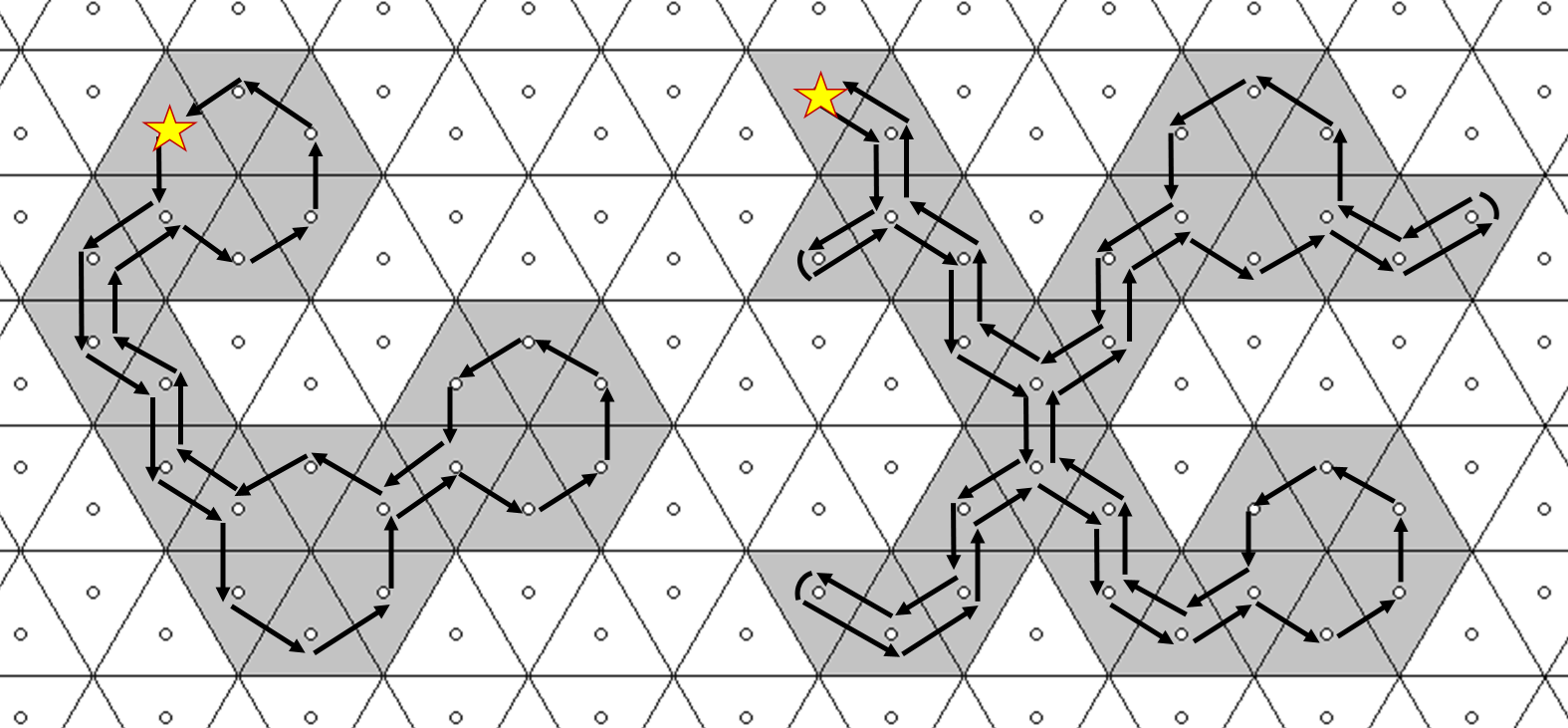}
\caption{Freeman code for the edge-adjacency in the triangular tiling, and two complexes, each with its canonical boun\-da\-ry path. Due to Algorithm \ref{alg:bound-tracing}, that path starts at the tile marked by a star. In the middle figure, this is an upright triangle of a regular complex, the boundary Freeman chain code begins as $(2,2,2,0,2,0,2,0,0,1,0,,0,0,\cdots )$. In the right figure, the complex is not regular, the boundary path starts at an inverted triangle which is an end tile, the boundary Freeman chain code begins as $(0,2,2,0,0,2,0,2,2,2,2,1,0,0,\cdots )$.}
\label{fig:Freeman-code-boundary}
\end{figure}

Now consider an object in a triangular tiling $\mathcal{M}$, i.e., a regular complex $\mathcal{C}\subset\mathcal{M}$. Then $\mathcal{C}$ is edge-\-adja\-cen\-cy-\-connected and has no end tile, and $|\mathcal{C}|$ is a simple polygon, so, $fr(|\mathcal{C}|)$ is a Jordan curve. Moreover, suppose that $\mathcal{C}$ has at least two tiles. The boundary path of $\mathcal{C}$ may repeat tiles, observe the examples in Figure \ref{fig:Freeman-code-boundary}. 

Moreover, let $\mathcal{N}$ be a finite set with $\mathcal{C}\subset\mathcal{N}\subset\mathcal{M}$ and such that the \textit{background tiles} from $(\mathcal{N}\setminus\mathcal{C})$ completely surround the complex $\mathcal{C}$. The adjacency graph $G(\mathcal{M})$ has the finite induced subgraphs $G(\mathcal{N})$ and $G(\mathcal{C})$, where each node has a valence at most three. Similar as known for other neigh\-bor\-hood graphs, we use the \textit{Freeman chain code} to represent paths in $G(\mathcal{N})$, see Figure \ref{fig:Freeman-code-boundary}. For a tile $T\in G(\mathcal{N})$, the movement direction from $T$ to any of its neighbors $T'$, is encoded by a number $f(T,T')\in \{ 0,1,2\}$. Any path has alternating upright and inverted triangles. If the type of the starting tile is known, a path can be reconstructed in a unique way from its Freeman chain code.

Boun\-da\-ry tracing begins by finding a first boundary tile. One way is to detect the leftmost tile $T_0$ of the top row of $\mathcal{C}$, as in Algorithm \ref{alg:bound-tracing}. Clearly $T_0$ belongs to $\mathcal{B}(\mathcal{C})$. This can be performed by scanning $\mathcal{N}$ or $G(\mathcal{N})$ on horizontal rows from the left to the right. Then $T_0$ has a left neighbor in the same row that belongs to the background: if $T_0$ is an upright triangle and $f(T_0,q)=1$, or if  $T_0$ is an inverted triangle and $f(T_0,q)=2$, then $q\not\in\mathcal{C}$. For each last found boundary tile $T_n$, Algorithm \ref{alg:bound-tracing} determines the Freeman code $b=f(T_n, T_{n-1})$ and then inspects the tile $T'$ that lies in direction $((b+1)\!\!\mod 3)$: if $T'$ belongs to $\mathcal{C}$ it is the next boundary tile found. Otherwise, the tile $T'$ positioned in direction $((b+2)\!\!\mod 3)$ is examined; it is the next boundary tile if it lies in $\mathcal{C}$. If not, the tile $T'$ lying in direction $((b+3)\!\!\mod 3)$ belongs to $\mathcal{C}$ and is the next boundary tile. The last case only occurs if $T_n$ is an end tile of $\mathcal{C}$. This process can be simplified for a regular complex which has no end tile. 

Algorithm \ref{alg:bound-tracing} determines a uniquely defined boun\-da\-ry path $\beta(\mathcal{C}) =$ $(T_0,$ $T_1,$ $T_2, \cdots ,$ $T_k)$, such that if boun\-da\-ry tracing would continue, the next tiles found would be, again, $T_{k+1}=T_0, T_{k+2}=T_1$. Step 2 will again found $T_0$ when boun\-da\-ry tracing is complete, but this can also happen if the boun\-da\-ry path touches itself at $T_0$; the end condition test distinguishes between the two si\-tua\-tions. Since each Freeman code from Figure \ref{fig:Freeman-code-boundary} generates a counterclockwise ordering of the neigbors around a tile, the resulting boundary path follows the boundary counterclockwise. Figure \ref{fig:Freeman-code-boundary} also illustrates the properties of Lemma \ref{lemma:canonic-boundary-leaves-C-onLeft}, which are evident from the performance of Algorithm \ref{alg:bound-tracing}. Appropriate changes in Step 1 can modify Algorithm \ref{alg:bound-tracing} to start with another boundary tile. The boundary path resulting from such modification, is a shifted version of the cyclic sequence $\beta(\mathcal{C})$ obtained from Algorithm \ref{alg:bound-tracing}. This justifies the following definition.

\begin{algorithm}
\begin{algorithmic}[1]

\Require A finite subset $\mathcal{N}$ of the triangular tiling, a complex $\mathcal{C}\subset \mathcal{N}$ that is surrounded by tiles from the background $(\mathcal{N}\setminus\mathcal{C})$.

\Ensure List $\beta(\mathcal{C})$, a boun\-da\-ry path of $\mathcal{C}$.

\hrule
\Statex

\State \textbf{Step 0:} Find the leftmost tile $T_0$ of the top row of $\mathcal{C}$.

\State \textbf{Step 1:} Initialize the list $\beta(\mathcal{C}):=(T_0)$. Set $b:=1$ if $T_0$ is an upright triangle, set $b:=2$ if $T_0$ is an inverted triangle. Then perform iteratively $b:=(b+1)\!\!\mod 3$. For each $b$, check whether the tile $T'$ that satisfies $f(T_0,T')=b$, belongs to $\mathcal{C}$. While this is false, $b$ is augmented to continue searching for a tile of $\mathcal{C}$. If $\, T'\in \mathcal{C}$, add $T'$ to $\beta(\mathcal{C})$ which becomes $\beta(\mathcal{C})= (T_0, T_1)$.

\State \textbf{Step 2:} Determine $b:=f(T_n,T_{n-1})$ for the current list $\beta(\mathcal{C})=$ 
$(T_0,$ $T_1,$ $T_2, \cdots ,$ $T_n)$. Then perform iteratively $b:=(b+1)\!\!\mod 3$. For each $b$ analyse whether the tile $T'$ satisfying $f(T_n,T')=b$ belongs to $\mathcal{C}$. While $\, T'\not\in \mathcal{C}$, $b$ is increased to continue searching for a tile of $\mathcal{C}$. If $\, T'\in \mathcal{C}$, proceed to the End Condition Test.

\State \textbf{End Condition Test:}

\If {$T'\neq T_0$} add $\, T'$ to $\beta(\mathcal{C})$, then go to Step 2.

\Else $\, $ (now $T'= T_0$) add $\, T'$ to $\beta(\mathcal{C})$. Then run Step 2 with the current list $\beta(\mathcal{C})$ only to obtain a new next boun\-da\-ry tile $\, T'\in\mathcal{C}$. Then,

\If {$T'=T_1$} remove the last tile from $\  \beta(\mathcal{C})$, then STOP.

\Else $\, $ add $\, T'$ to $\beta(\mathcal{C})$, $\  $ then go to Step 2.

\EndIf
\EndIf
\end{algorithmic}
\caption{Boun\-da\-ry tracing for a regular complex $\mathcal{C}$ with at least two tiles, in a triangular tiling.}
\label{alg:bound-tracing}
\end{algorithm}

\begin{definition}
Let $\mathcal{C}$ be a regular complex with at least two tiles, in a triangular tiling. A boun\-da\-ry path $b=(t_0, t_1, \cdots , t_k)$ of $\mathcal{C}$ which, up to some shift of the cyclic sequence $b$, coincides with the sequence $\beta(\mathcal{C})$ determined by the boun\-da\-ry tracing Algorithm \ref{alg:bound-tracing}, is called the \textbf{\textit{canonical boun\-da\-ry path}} of $\mathcal{C}$.
\label{def:canonical-bound}
\end{definition}

The canonical boun\-da\-ry path follows the boundary counterclockwise and is unique up to shifts of the cyclic sequence $\beta(\mathcal{C})$. Here we are interested in regular complexes, which clearly satisfy the assumptions of Definition \ref{def:canonical-bound}.

\begin{figure}
\centering
\includegraphics[height=2.2cm]{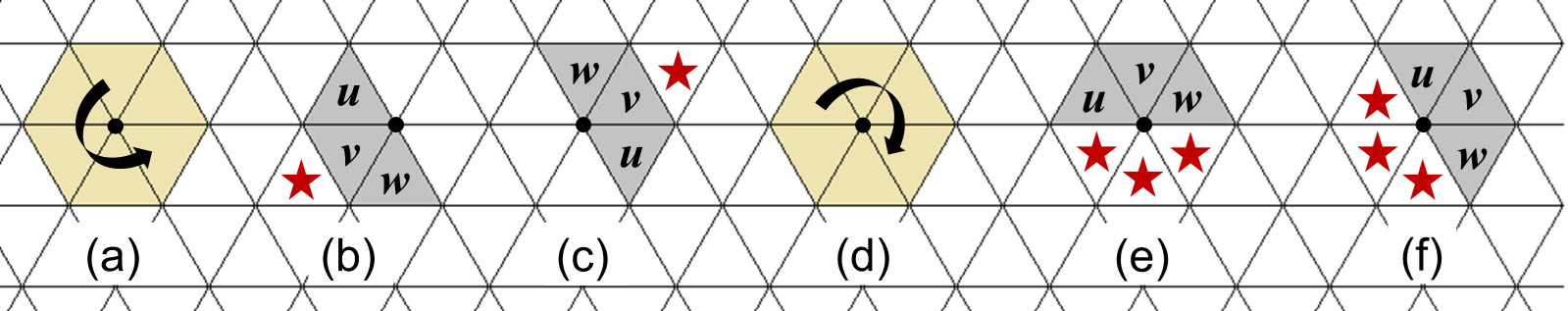}
\caption{Possible situations of triples $(u,v,w)$ in $\beta(\mathcal{C})$, used in the proof of Lemma \ref{lemma:canonic-boundary-leaves-C-onLeft}, starting with $u$ being any tile in the hexagon shown in (a) or (d). The centre point $c_i$ of $v$ is a convex vertex of $\gamma (\beta(\mathcal{C}))$ if $(u,v,w)$ follows an order as in (a), two rotational instances of $(u,v,w)$ within the hexagon are shown in (b),(c), its central vertex is drawn, the tile marked by a star must lie in the background. The centre $c_i$ of $v$ is a concave vertex of $\gamma (\beta(\mathcal{C}))$ if $(u,v,w)$ are ordered as in (d); (e) and (f) present two rotational instances of $(u,v,w)$ within that hexagon, at least one of the tiles marked by a star, must lie in the background.}
\label{fig:triples-bound-convex-concave}
\end{figure}

Given a path $b= (t_0, t_1, \cdots , t_k)$ in the triangular tiling, denote by $c_i$ the centre point of $t_i$, $i\in\{ 0,1, \cdots , k\}$, and consider the curve $\gamma (b) = \overline{c_0c_1} \cup $ $\overline{c_1c_2}\cup \cdots \cup $ $\overline{c_{k-1}c_k}$. Then, for any consecutive tiles $t_{i-1}$, $t_{i}$, $t_{i+1}$ in $b$, $c_i$ is not a linear point of $\gamma (b)$: supposing that $t_{i-1} \neq t_{i+1}$, $\overrightarrow{c_{i-1}c_i}$ and $\overrightarrow{c_ic_{i+1}}$ form an angle of $120^{\circ}$ or $240^{\circ}$, that is, $c_i$ is a concave or convex vertex of $\gamma (b)$. For the following definition, see Figure \ref{fig:triples-bound-convex-concave}.

\begin{definition}
Let $\mathcal{C}$ be a regular complex with at least two tiles, in a trian\-gu\-lar tiling, and $b=(t_0, t_1, \cdots , t_k)$ any boun\-da\-ry path of $\mathcal{C}$.

\noindent (1) If $c_i$, $i\in\{ 0,1,\cdots , k\}$, denotes the centre point (i.e., the incircle centre) of the triangle $t_i$, the curve $\gamma (b) =\overline{c_0c_1} \cup $ $\overline{c_1c_2}\cup \cdots \cup $ $\overline{c_{k-1}c_k} \cup \overline{c_kc_0}$ is called the \textbf{poly\-gonal curve of centre points} of $b$.

\noindent (2) We say that $b$ \textbf{always leaves the background to the right} if each triple 
$(t_{i-1} , t_{i} , t_{i+1})$ (with index calculus modulo (i+1)) of consecutive tiles in $b$ satisfies exactly one of the following two options:

\noindent $\, \bullet\, $ $c_i$ is a convex vertex of $\gamma (b)$. Taking the Freeman code $f= f(t_{i} , t_{i+1})$, the neighbor of $t_{i}$ that lies in direction $((f-1)\!\!\mod 3)$ does not belong to $\mathcal{C}$.

\noindent $\, \bullet\, $ $c_i$ is a concave vertex of $\gamma (b)$. At least one of the other three tiles that also belong to the hexagon composed of six tiles containing $t_{i-1} , t_{i} , t_{i+1}$ does not belong to $\mathcal{C}$.

\label{def:boundary-leaves-backgr-onRight}
\end{definition}

In particular, the curve $\gamma (\beta(\mathcal{C}))$ for the canonical boundary path $\beta(\mathcal{C})$ of a regular complex is traced counterclockwise, and $\beta(\mathcal{C})$ always leaves the background to the right. Clearly, then $\beta(\mathcal{C})$ always leaves the complex $\mathcal{C}$ to the left or on itself. This important property even characterizes the canonical boundary path, as the following lemma shows. 

\begin{lemma}
Let $\mathcal{C}$ be a regular complex $\mathcal{C}$ with at least two tiles in a triangular tiling $\mathcal{M}$, $\beta(\mathcal{C}) = (\beta_0 , \beta_1 , \cdots , \beta_k)$ its canonical boun\-da\-ry path, and $\gamma (\beta(\mathcal{C})) = \overline{c_0c_1} \cup \overline{c_1c_2}\cup \cdots \cup \overline{c_{k-1}c_k} \cup \overline{c_kc_0}$ the polygonal curve of centre points of the tiles of $\beta(\mathcal{C})$.

\noindent (1) $\beta(\mathcal{C})$ is regular: $\beta_{i-1}\neq \beta_{i+1}$ for all $1\leq i\leq k-1$, $\beta_1\neq \beta_k$, $\beta_{k-1}\neq \beta_0$.

\noindent (2) $\beta(\mathcal{C})$ always leaves the background to the right.

\noindent (3) Any boundary path $b$ of $\mathcal{C}$ that always leaves the background to the right coincides with the canonical boundary path $\beta(\mathcal{C})$.

\label{lemma:canonic-boundary-leaves-C-onLeft}
\end{lemma}

\noindent Proof:
\smallskip

\noindent (1) follows from Algorithm \ref{alg:bound-tracing} since $\beta(\mathcal{C})$ has no end tiles.

\noindent (2) Let $\beta_{i-1} , \beta_{i} , \beta_{i+1}$ be consecutive tiles in $\beta(\mathcal{C})$, denote $\gamma =\gamma (\beta(\mathcal{C}))$. By (1), $\beta_{i-1}\neq \beta_{i+1}$, hence the part $\overline{c_{i-1}c_i} \cup \overline{c_ic_{i+1}}$ of $\gamma $ forms an interior angle of $120^{\circ}$, which is equivalent to that $c_i$ is a convex vertex of $\gamma$, or forms an interior angle of $240^{\circ}$ being equivalent to that $c_i$ is a concave vertex of $\gamma $. Figure \ref{fig:triples-bound-convex-concave} shows possible situations for $u= \beta_{i-1} , v= \beta_{i} , w= \beta_{i+1}$.

That $c_i$ is convex corresponds to one of six possible situations as those shown in Figure \ref{fig:triples-bound-convex-concave}(b)(c), which are rotated versions of the same hexagon of six tiles containing $u, v, w$. Without loss of generality, we consider only case (b). The triple $(u,v,w)$ is part of $\beta(\mathcal{C})$ and hence is found by Algorithm \ref{alg:bound-tracing}. Starting Step 2 with $T_n=v$, Algorithm \ref{alg:bound-tracing} determines the Freeman code $f(v,u)$ and augments it by 1 (modulo 3), giving a direction to reach tile $T'$ that is precisely the neighbor of $v= \beta_i$ marked by a star in \ref{fig:triples-bound-convex-concave}(b). If $T'\in \mathcal{C}$, $T'$ is the next boundary tile found, then $c_i$ is not a convex vertex of $\gamma $, which contradicts the hypothesis, hence $T'$ lies in the background.

That $c_i$ is concave corresponds to one of six possibilities as those shown in Figure \ref{fig:triples-bound-convex-concave}(e)(f), which are rotated versions of the same hexagon of six tiles containing $u, v, w$, denote by $s$ the central vertex of this hexagon. Without loss of generality, we treat only case (e). Algorithm \ref{alg:bound-tracing} finds the triple $(u,v,w)$, $fr(|\mathcal{C}|)$ passes on the right side of the tiles $u= \beta_{i-1}, v= \beta_{i}, w= \beta_{i+1}$. Since $\beta(\mathcal{C})$ is regular, $\beta_{i-2}$ lies below $u$ or is the left neighbor of $u$. Also, $\beta_{i+2}$ lies below $w$ or is the right neighbor of $w$. In consequence, $s\in fr(|\mathcal{C}|)$. Assuming that all tiles marked by a star in Figure \ref{fig:triples-bound-convex-concave}(e) belong to $\mathcal{C}$ implies that $p\not\in fr(|\mathcal{C}|)$, which is a contradiction. In consequence, at least one of these tiles lies in the background. 

\noindent (3) Let $b= (t_0, t_1, \cdots , t_k)$ be a boundary path of $\mathcal{C}$ which always leaves the background to the right, and $\gamma (b) = \overline{c_0c_1} \cup \overline{c_1c_2}\cup \cdots \cup \overline{c_{k-1}c_k} \cup \overline{c_kc_0}$ the polygonal curve of centre points of the tiles of $b$. For any $t_i$, $0\leq i\leq k$, the triple $(t_{i-1} , t_i , t_{i+1})$ (with index calculus modulo $(k-1)$) satisfies the hypothesis.

If $c_i$ is a convex vertex of $\gamma (b)$, up to rotations of the hexagon that contains $u=t_{i-1} , v=t_i , w= t_{i+1}$, the situation is like in Figure \ref{fig:triples-bound-convex-concave}(b), where the tile $T$ marked by a star does not belong to $\mathcal{C}$. Algorithm \ref{alg:bound-tracing} finds the tile $v=t_i$ from the previous one $t_{i-1}$, then determines the Freeman direction $f= f(t_i, t_{i-1})$ and inspects the tile in direction $f'=(f+1 \mod 3)$, but this tile is exactly $T$. Since $T\not\in \mathcal{C}$, Algorithm \ref{alg:bound-tracing} determines $f'' = (f'+1 \mod 3)$ to examine the tile $t_{i+1}$ that lies in direction $f''$. Since $t_{i+1}\in \mathcal{C}$ it is the next boundary tile found for the sequence $\beta(\mathcal{C})$.

If $c_i$ is a concave vertex of $\gamma (b)$, up to rotations of the hexagon that contains $u=t_{i-1} , v=t_i , w= t_{i+1}$, the situation is like in Figure \ref{fig:triples-bound-convex-concave}(e). Algorithm \ref{alg:bound-tracing} finds $v=t_i$ from $t_{i-1}$, then determines the Freeman direction $f= f(t_i, t_{i-1})$ and inspects the tile in direction $f'=(f+1 \mod 3)$, but this tile coincides with $w= t_{i+1}$, which belongs to $\mathcal{C}$ and hence is the next tile found for $\beta(\mathcal{C})$.

In consequence, the cyclic sequences $b$ and $\beta(\mathcal{C})$ coincide.

\qed
\medskip

The last lemma states that the canonical boundary path is the unique boun\-da\-ry path of $\mathcal{C}$ which always leaves the background to the right. Although this paper deals only with regular complexes, we mention that Algorithm \ref{alg:bound-tracing} can be applied to more general complexes, even having end tiles, where it also generates a uniquely defined boundary path with pro\-per\-ties similar as those in Lemma \ref{lemma:canonic-boundary-leaves-C-onLeft}.

Since Algorithm \ref{alg:bound-tracing} examines each boundary tile together with at most its three neighbors, it has linear time complexity depending of the number of boundary tiles of the complex $\mathcal{C}$. Although $\mathcal{C}$ may coincide with its boundary if $\mathcal{C}$ is a ``thin" complex, for most practical purposes the boundary has far fewer tiles than
the complex itself.


\section{Minimum perimeter polygon for regular complexes in triangular tilings}
\label{sect:MPP-object-regular-complex}


\begin{figure}
\centering
\includegraphics[height=3.5cm]{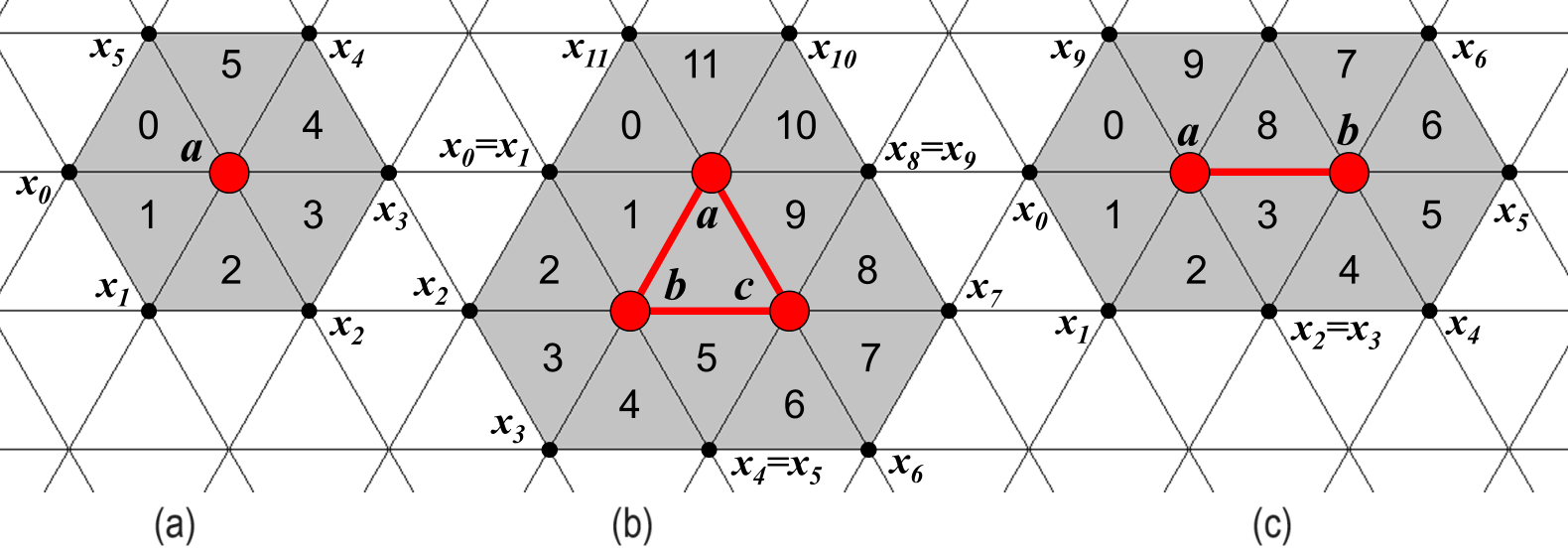}
\caption{For three regular complexes (shaded grey), their core frontiers are drawn in red, each number $k$ within a tile indicates $\beta_k$ in the canonical boundary path. For each complex, the core is connected, all core vertices are convex, and the core coincides with the MPP.}
\label{fig:example1}
\end{figure}

Recall that a regular complex $\mathcal{C}$ is edge-\-adja\-cen\-cy-\-connec\-ted, has no end tile, and $|\mathcal{C}|$ is a simple polygon. 
If $\mathcal{C}$ has at least two tiles, it is easy to see that the edge-\-connec\-ted\-ness and the lack of end tiles imply that $\mathcal{C}$ has at least six tiles forming a hexagon. Figure \ref{fig:example1} shows some special regular complexes, which will be used in Example 1.

\noindent \textbf{--} The complex in Figure \ref{fig:example1}(a), which consists of six tiles forming a hexagon, will be denoted by $\mathcal{C}_{Hex}$, it is the smallest regular complex with at least two tiles. The core of $\mathcal{C}_{Hex}$ is the common vertex of the six tiles, it coincides with its MPP and is a polygon degenerated to a single point.

\noindent \textbf{--} The complex in Figure \ref{fig:example1}(c) with 10 tiles forming two over\-lapping hexa\-gons is the smallest regular complex with more than 6 tiles. Its core is a straight line segment that coincides with the MPP of $\mathcal{C}$.

\noindent \textbf{--} The complex in Figure \ref{fig:example1}(b) consists of 13 tiles forming three overlapping hexagons. Its core coincides with its MPP, which is a triangle whose vertices are given as the central vertices of the three hexagons.

For a complex $\mathcal{C}$, an open simple path in $\mathcal{C}$ is a path $(T_1, T_2, \cdots , T_k)$ of tiles of $\mathcal{C}$ such that for each $i\in \{ 2,3, \cdots , k-1\}$, exactly two neighbors of $T_i$ belong to this path, but no other neighbor of $T_i$ belongs to $\mathcal{C}$. Intuitively, this is a thin part of $\mathcal{C}$. 

\begin{lemma}
Any regular complex $\mathcal{C}$ with at least two tiles in a triangular tiling is a union of hexagons of type $\mathcal{C}_{Hex}$ and, eventually, simple open paths in $\mathcal{C}$ connecting parts of $\mathcal{C}$ that are unions of hexagons $\mathcal{C}_{Hex}$.
\label{lemma:regular-complex-hex}
\end{lemma}

\noindent Proof:
The complex $\mathcal{C}_{Hex}$, which consists of six tiles forming a hexagon, is the smallest regular complex with at least two tiles. Any larger regular complex $\mathcal{C}$ contains $\mathcal{C}_{Hex}$ and an additional tile $T_7$ that is (edge-) adjacent to some tile of $\mathcal{C} _{Hex}$. Since $T_7$ is not an end tile and $\mathcal{C}$ is connected, another new tile $T_8$ is needed. $T_8$ is adjacent to $T_7$, but not necessarily to the tiles of $\mathcal{C} _{Hex}$. This process may be continued in two ways: by adding new tiles which create the next hexagon of type $\mathcal{C} _{Hex}$, or by generating a simple open path of arbitrary length. At each end of every open simple path in $\mathcal{C}$, a hexagon as $\mathcal{C} _{Hex}$ (or a union of such hexagons) is required to avoid the presence of end tiles.

\qed
\medskip

The regular complex in Figure \ref{fig:Freeman-code-boundary} is the union of three hexagons of type $\mathcal{C} _{Hex}$ and a simple open path $\mathcal{C}$, which connects two hexagons. Figure \ref{fig:Introd-example-regular-complex-MPP} presents another regular complex with its MPP.
These examples show that the core of a regular complex $\mathcal{C}$ can be disconnected. Each hexagon $\mathcal{C} _{Hex}$ in $\mathcal{C}$ which does not overlap with other such hexagons contributes an isolated point to \textit{core}$(\mathcal{C})$. Overlapping hexagons of type $\mathcal{C} _{Hex}$ in $\mathcal{C}$ generate parts of the core that contain edges or triangle tiles. 

Due to Theorem 4.1 of \cite{SklanskyKib1976}, for any points $a,b$ in a polygon $B$ there is a unique polygonal shortest path, called \textit{geodesic} in $B$, which connects $a$ and $b$. For two polygons $A\subseteq B$, $A$ is called \textit{convex relative} to $B$ if $A$ contains each geodesic in $B$ that connects points $a,b$ from $A$. The \textbf{\textit{relative convex hull}} of $A$ with respect to $B$ is the intersection of all sets containing $A$ that are convex relative to $B$. 

By Theorem 3.10 and Corollary 4.3 of \cite{SklanskyKib1976}, if $\mathcal{C}$ is a normal complex in a polygonal tiling such that the union of any two adjacent tiles forms a convex polygon, then the MPP of $\mathcal{C}$ exists and is unique. Moreover, the MPP then coincides with the relative convex hull of a set $S$ with respect to $|\mathcal{C}|$, where $S$ is the set of all core vertices and cut points of $\mathcal{C}$, $S$ is called the \textit{spread} of $\mathcal{C}$ in \cite{SklanskyKib1976}. A regular complex is normal but has no cut points. The special regular complex $\mathcal{C}$ consisting of a unique tile $T$ has an empty core, and, any point of $T$ is a polygon that satisfies the definition of the MPP. Taking caution of this, the following is a consequence from the facts shown in \cite{SklanskyKib1976}:

\begin{lemma} 
If $\mathcal{C}$ is a regular complex with at least two tiles in a triangular tiling, then $\mathcal{C}$ has a unique MPP that coincides with the relative convex hull of $\hbox{\textit{core}}(\mathcal{C})$ with respect to $|\mathcal{C}|$. Specially, the MPP contains $\hbox{\textit{core}}(\mathcal{C})$ and is contained in $|\mathcal{C}|$. 
\label{lemma:known-for-regular-complex}
\end{lemma}

Our MPP algorithm presented in Section \ref{sect:MPP-Algorithm} uses the canonical boun\-da\-ry path $\beta(\mathcal{C})$ as input data, where the following property is important.

\begin{lemma}
If $\mathcal{C}$ is a regular complex with at least six tiles in a triangular tiling $\mathcal{M}$ and $\beta(\mathcal{C}) = (\beta_0 , \beta_1 , \cdots , \beta_k)$ is its canonical boun\-da\-ry path, then the frontier curve of the MPP of $\mathcal{C}$ is the curve with the shortest length among all polygonal curves that visit all tiles $\beta_0 , \beta_1 , \cdots , \beta_k$ in the order according to the list $\beta(\mathcal{C})$.
\label{lemma:MPP-shortest-curve-through-canon-bound-path}
\end{lemma}

\noindent Proof:
Denote by $P$ the MPP of $\mathcal{C}$, its frontier $\gamma$ is a weakly simple polygonal curve. By Lemma \ref{lemma:known-for-regular-complex}, $P$ is the relative convex hull of \textit{core}$(\mathcal{C} )$ with respect to $|\mathcal{C}|$, in particular, \textit{core}$(\mathcal{C} )\subset P \subset |\mathcal{C}|$. Since $P$ is a preimage of $\mathcal{C}$, $P$ reaches all tiles of $\mathcal{C}$, in particular all boundary tiles. When $\gamma$ is traced  counterclockwise, it takes all components of \textit{core}$(\mathcal{C} )$ into $P$ and leaves them on its left side (or on itself). By Lemma \ref{lemma:canonic-boundary-leaves-C-onLeft}, $\beta(\mathcal{C})$ is the unique boundary path that always leaves the complex tiles always to the left (or on itself) and the background strictly to the right. Therefore, $\gamma$ passes through all boundary tiles according to their order in $\beta(\mathcal{C})$. Moreover, $\gamma$ is the shortest such curve, since any polygonal curve shorter than $\gamma$ but also passing through all boundary tiles in the order as in $\beta(\mathcal{C})$ would be the frontier of a polygon that is a preimage of $\mathcal{C}$ and has a shorter perimeter than $P$, contradicting the assumption that $P$ is the MPP of $\mathcal{C}$. 

\qed
\medskip

The following property is also useful for our MPP algorithm.

\begin{lemma}
Let $\mathcal{C}$ be a regular complex with at least six tiles in a triangular tiling $\mathcal{M}$.

\noindent (1) The leftmost and the rightmost tiles of the top row of $\mathcal{C}$ are upright triangles. 

\noindent (2) The leftmost and the rightmost tiles of the bottom row of $\mathcal{C}$ are inverted triangles.
\label{lemma:regular-complex-starttile}
\end{lemma}

\noindent Proof: Let $T$ be the leftmost tile of the top row of $\mathcal{C}$. By the hypothesis, $T$ is adjacent to another tile of $\mathcal{C}$. Supposing that $T$ is an inverted triangle, only its neighbor to the right can belong to $\mathcal{C}$. Hence $T$ is an end tile of $\mathcal{C}$, which contradicts the regularity of $\mathcal{C}$. In consequence, $T$ is an upright triangle. Analogously, if the rightmost tile $T$ of the top row of $\mathcal{C}$ is an inverted triangle, it is an end tile since only its neighbor to the left belongs to $\mathcal{C}$, this contradiction implies that $T$ is an upright triangle.
Similarly, if $T$ is the leftmost or rightmost tile in the bottom row of $\mathcal{C}$, then $T$ is an inverted triangle because $T$, assumed to be an upright triangle, is an end tile of $\mathcal{C}$.

\qed
\smallskip


\section{An algorithm to determine the minimal perimeter polygon for regular complexes in triangular tilings}
\label{sect:MPP-Algorithm}


Algorithm \ref{alg:MPP-Alg-triangular} takes as input data the canonical boun\-da\-ry path $\beta (\mathcal{C} )=$ $(\beta_0,$ $\beta_1,$ $\cdots ,$ $\beta_t)$ where $\beta_0$ is the leftmost tile $T_0$ of the top row of $\mathcal{C}$. If the canonical boundary path was obtained by another algorithm, cyclic shifting in the list $\beta (\mathcal{C})$ achieves that $T_0 = \beta_0$. We suppose that the list $\beta (\mathcal{C} )$ provides information about the vertices of each tile $\beta_i$, which facilitates the computation of the endpoints $x_i,y_i$ of each edge $e_i= \beta_i\cap \beta_{i+1}$.

\begin{algorithm}
\begin{algorithmic}[1]

\Require canonical boun\-da\-ry path $\beta (\mathcal{C} )= (\beta_0, \beta_1,\cdots , \beta_t)$ (counterclockwise tracing) where $\beta_0$ is the leftmost tile in the top row of $\mathcal{C}$.

\Ensure List \textit{MPP} of MPP vertices.

\hrule

\Statex

\State Determine the lower right corner point $m$ of $\beta_0$.

\State Add $m_1=m$ to the list \textit{MPP}, set $i:=0$ , $n:=1$.

\State \textbf{if} $i> t$ \textbf{then} STOP \textbf{endif}

\State Determine the endpoints $x_i$ (right) and $y_i$ (left) of the edge $e_i= \beta_i\cap\beta_{i+1}$.

\If {($m\neq x_i$ and $m\neq y_i$)}

\State $p:=x_i$ (initial right cone border $\overrightarrow{mp}$)

\State $q:=y_i$ (initial left cone border $\overrightarrow{mq}$)

\State $k:=i$

\While {($m,y_k,x_k$ form a right turn or are collinear, and $x_k,y_k$ do not

\noindent\hskip0.5cm both lie on the same side strictly outside the cone)}

\State Update right border:

\noindent\hskip1cm \textbf{if} ($m,p,x_k$ form a left turn or are collinear) \textbf{then}

\noindent\hskip1cm ($p:=x_k$ and $z_p:=k$) \textbf{endif}

\State Update left border:

\noindent\hskip1cm \textbf{if} ($m,q,y_k$ form a right turn or are collinear) \textbf{then}

\noindent\hskip1cm ($q:=y_k$ and $z_q:=k$) \textbf{endif}

\State $k:=k+1$  (continue with the same cone)

\State \textbf{if} $k> t$ \textbf{then} STOP, \textbf{endif}

\State Determine the endpoints $x_k$ (right) and $y_k$ (left) of $e_k= \beta_k\cap\beta_{k+1}$.

\EndWhile

\State (A new MPP vertex is found:)

\State \textbf{if} ($m,p,x_k$ form a right turn or are collinear) \textbf{then}

\noindent\hskip0.5cm ($m:= p$ and $z:=z_p$) \textbf{endif}

\State \textbf{if} ($m,p,x_k$ form a left turn) \textbf{then}

\noindent\hskip0.5cm ($m:= q$ and $z:=z_q$) \textbf{endif}

\State $n:=n+1$, add the point $m_n=m$ to the list \textit{MPP}.

\State $i:= z+1$ (continue searching with $\beta_{z+1}$)

\State Go to Line 3.

\EndIf

\State $i:=i+1$

\State Go to Line 3.
\end{algorithmic}
\caption{Determination of the ordered list of MPP vertices for a regular complex $\mathcal{C}$ with at least six tiles in the tiling of equilateral triangles.}
\label{alg:MPP-Alg-triangular}
\end{algorithm}

Algorithm \ref{alg:MPP-Alg-triangular} constructs a polygon $P$ whose all vertices are vertices of the tiles of $\beta(\mathcal{C})$. In Line 1, the first polygon vertex $m_1$ is determined; $m_1$ is an MPP vertex due to Lemma \ref{lemma:MPP-initial-vertex}. In Line 4, the endpoints $x_i, y_i$ of $e_i= \beta_i\cap\beta_{i+1}$ are determined such that $x_i$ lies to the right of $\overrightarrow{c_ic_{i+1}}$ if $c_i$ denotes the centre point of $\beta_i$, and $y_i$ lies to the left of $\overrightarrow{c_ic_{i+1}}$, $x_i\in fr(|\mathcal{C} |)$ by Lemma \ref{lemma:canonic-boundary-leaves-C-onLeft}, the same applies to $x_k, y_k$ in Line 14.

From each polygon vertex $m=m_n$, $n\geq 1$, provided by some $\beta_z\in \beta(\mathcal{C})$, the next polygon vertex is found with the help of a cone of visibility through the forthcoming tiles. The cone rooted at $m$ is initialized in Lines 6-7 by its right border $\overrightarrow{mp}$ and left border $\overrightarrow{mq}$ with $p=x_i$, $q=y_i$, which are the right and left endpoints of $e_i= \beta_i \cap \beta_{i+1}$ where $i$ is the first index larger than $z$ such that $x_i$ and $y_i$ are distinct from $m$. The cone is updated in Lines 10-11 after inspecting each $\beta_k$: $p=x_k$ is updated if $x_k$ restricts or confirms the right border, $q=y_k$ is updated if $y_k$ restricts or confirms the left border. The variable $p$ always is the last point that defined the right cone border, it was provided as some $x_k$, the variable $z_p$ stores the index of the tile $\beta_k$ in the list $\beta(\mathcal{C})$. Similarly, $q$ is the last point defining the left cone border provided as some $y_k$, $z_q$ stores this index $k$. When a new polygon vertex is found, it is given as the point $p$ or $q$ which originally was provided by a tile with an index stored as $z_p$ or $z_q$ and transferred in Line 17 or 18 to the variable $z$. The search for the next polygon vertex starts just with the tile having the next index, which is guaranteed by Line 20.

The condition of Line 9 requires that $m,y_k,x_k$ form a right turn or are collinear and that $x_k, y_k$ do not both lie on the same side strictly outside the cone. Only in this case, the cone of visibility can be extended up to the current tile $\beta_k$ and eventually is updated. It is important that in Line 10, $x_k$ is ignored if it lies strictly outside the cone on the right because then $x_k$ must not be used to update the right cone border, since the straight line $\overline{mx_k}$ leaves the boundary. The same occurs in Line 11: $y_k$ is ignored if it lies strictly outside the cone on the left since $\overline{my_k}$ leaves the boundary. Lines 9-11 guarantee that $(m,p,q)$ well-defines the cone and that $\overline{mp}$ and $\overline{mq}$ always belong to the union of corresponding tiles from $\beta_z$ up to the tile which provides $p$ or $q$.

A new polygon vertex is found if the while condition in Line 9 is not satisfied, then the boundary performs at $\beta_k$ an essential movement to the left or to the right, which allows to find a new polygon vertex in Lines 17-19, respectively, as the point $q$ defining the left cone border, or the point $p$ determining the right cone border. We affirm that the polygon found by Algorithm \ref{alg:MPP-Alg-triangular} coincides with the MPP of $\mathcal{C}$ for any regular complex $\mathcal{C}$ with at least six tiles. 

\medskip\noindent \textbf{Example 1:} We apply Algorithm \ref{alg:MPP-Alg-triangular} to the complexes (a),(b),(c) of Figure \ref{fig:example1}.
 
\noindent \textbf{(a)} $\beta (\mathcal{C} )= (\beta_0 ,\beta_1 ,\cdots , \beta_5 )$ is the input path, Lines 1-2 yield $m=a =y_0$ and the initial list \textit{MPP}$=(a)$. Since $m=a= y_0 = y_1 = y_2 = y_3 = y_4 = y_5$, $i$ is stepwise augmented, but the condition of Line 5 is never satisfied, we get $i=6$ and arrive to Line 3 with $6>5$, which stops the algorithm. It returns the final list \textit{MPP}$=(a)$ that represents the MPP, which is a polygon degenerated to a single point.

\smallskip\noindent \textbf{(b)} $\beta (\mathcal{C} )= (\beta_0 , \beta_1,\cdots , \beta_{11})$, we get $m=m_1= y_0 = a$ and the initial list \textit{MPP}$=(a)$. For $i=1$, $\beta_1$ gives an initial cone with $p=x_1$ ($\overrightarrow{ax_1}$ right border) and $q= y_1 =b$ ($\overrightarrow{ay_1}$ left border). Now $k=1$, the condition of Line 9 is trivially satisfied, updates in Lines 10-11 change nothing. Line 12 gives $k=2$ to go back to Line 9, its condition is fulfilled for $x_2$, $y_2=b$, hence $p=x_2$ is updated, $q=y_2=b$ is confirmed. We reach Line 9 with $k=3$, its condition is satisfied hence $p=x_3$, $q=y_3=b$, the cone has degenerated to a ray. For $k=4$, the condition of Line 9 is not valid since $(m,y_4,x_4)$ forms a left turn. Line 18 provides $m_2 =m =q= y_3 =b$ which is added to the list \textit{MPP}.

Restarting in Line 3 with $m=b$, $i=4$, $y_4=m$, hence the initial cone is given by $p=x_5$ (right border) and $q=y_5 =c$ (left border). The cone is restricted later by $x_7$ where it becomes a ray, and $x_8$ lying outside on the left of $\overrightarrow{my_8}$ defines the next MPP vertex $m_3=m=y_8 =c$ which is enlisted in \textit{MPP}. Restarting with $m=c$ and $i=9$, $p=x_9$ and $q= y_9 = a = y_{10} = y_{11}$ provide the initial cone whose right border is restricted by $x_{10}$ and then by $x_{11} \in \overrightarrow{my_{11}}$ where the cone has become a ray. With $i=12 >11$ the algorithm stops and returns the correct list \textit{MPP}$=(a,b,c)$.

\smallskip\noindent \textbf{(c)} $\beta (\mathcal{C} )= (\beta_0 , \beta_1,\cdots , \beta_9)$, $m=m_1= y_0 = a$, \textit{MPP}$=(a)$. Since $y_1 = y_2 = a$, we get $i=3$, $\beta_3$ provides an initial cone given by $p=x_3$ (right border) and $q= y_3 =b$ (left border). Now $k=3$, the condition of Line 9 is satisfied, updates in Lines 10-11 change nothing. Coming back to Line 9 with $k=4$, its condition is fulfilled for $x_4$, $y_4=b$, hence $p=x_4$ is updated (restricts the cone), $q=y_4=b$ (confirms the cone). The condition of Line 9 with $k=5$ is satisfied hence $p=x_5$ and $q=y_5=b$, the cone has become a ray. For $k=6$, the condition of Line 9 is not valid since $(m,y_6,x_6)$ forms a left turn. The condition of Line 17 is not satisfied, but Line 18 gives $m_2 =m =q= y_6 =b$, $m_2$ is added to the list \textit{MPP}. It is easy to see that the algorithm proceeds until returning the correct list \textit{MPP}$=(a,b)$ of the MPP being a straight line segment.

\rightline{$\vartriangle$}

\begin{figure}
\centering
\includegraphics[height=5.4cm]{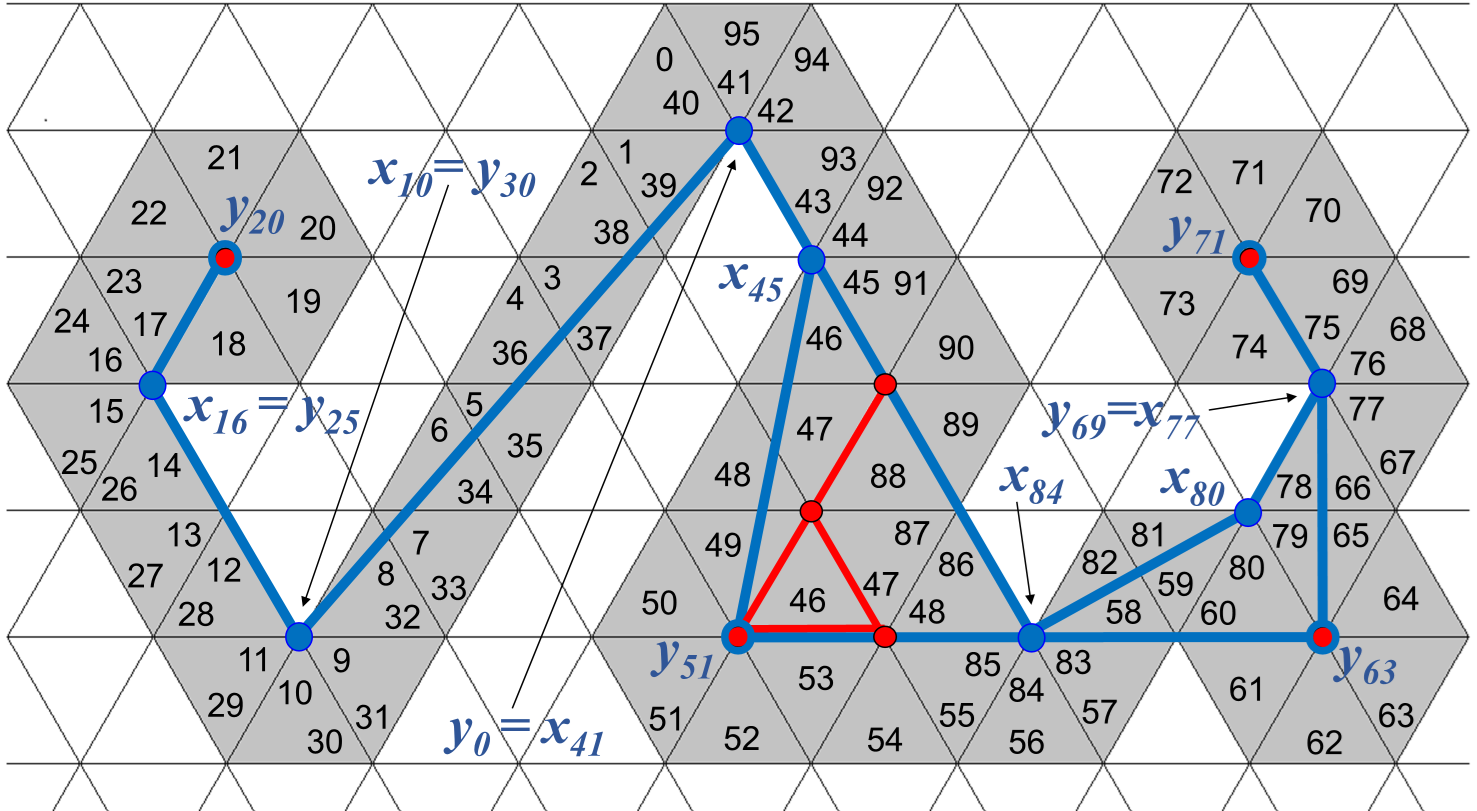}
\caption{In the regular complex shaded grey, each $\beta_k$ in the canonical boundary path is represented by the number $k$ shown in the tile, the core frontier is drawn in red. The MPP frontier depicted in blue is determined in Example 2.}
\label{fig:example2}
\end{figure}

\medskip\noindent \textbf{Example 2:} For the complex of Figure \ref{fig:example2}, the input list indicates $t=95$. Many boun\-da\-ry tiles are repeated in $\beta (\mathcal{C} )$, for example, $\beta_1 = \beta_{39}$, $\beta_{10}= \beta_{30}$, $\beta_{56} = \beta_{84}$. Algorithm \ref{alg:MPP-Alg-triangular} first finds $m_1 = m= y_0$. Then $\beta_1$ provides the initial cone with $p=x_1$ (right border) and $q=y_1 = y_2$ (left border). The points $y_3= y_4$, $y_5= y_6$, $y_7= y_8$ and $y_9$ all lie on the left cone border hence they confirm that border. But $x_2= x_3$, $x_4 = x_5$, $x_6 = x_7$ and $x_8= x_9$ define new restricted right borders. As result $p=x_9$ and $q=y_9$. Then $p= x_{10}= x_9$ confirms the right border, but $q= y_{10}$ gives a new restricting left border. For $\beta_{11}$, $x_{11} = x_{10}$ would confirm the right border, but the condition in Line 9 is violated since $(m,y_{11},x_{11})$ is a left turn, Line 17 yields $m_2= p= x_{10}$, $z=10$.

Restarting with $m=x_{10}$, $i=11$, $x_{11}= m$, $\beta_{12}$ provides the initial cone with $p= x_{12} = x_{13}$ (right border) and $q= y_{12}$ (left border). The right border is confirmed by $x_{14}= x_{15}$, but $y_{13}= y_{14}$ and $y_{15}$ restrict the cone by updating the left border, resulting in $p= x_{15}$ and $q= y_{15}$. The next updating gives $p= x_{16} = x_{15}$ and $q= y_{16}\in \overrightarrow{mx_{16}}$, the cone has become a ray. For $\beta_{17}$, $x_{17} = x_{16}$ but the condition in Line 9 is violated: $(m,y_{17},x_{17})$ forms a left turn, Line 17 results in $m_3= p= x_{16}$, $z=16$.

For $i=17$, $m= x_{16}= x_{17}$, $p= x_{18}$ and $q= y_{18} = y_{19} = y_{20}$ determine the initial cone. Its right border is restricted by $p= x_{19}$ and then by $p= x_{20}$ which lies on the left border, the cone becomes a ray. For $\beta_{21}$, $y_{21} = y_{20}$ would confirm the left border, but $(m,y_{21},x_{21})$ is a left turn which violates the condition of Line 9, Line 18 gives $m_4= q= y_{20}$, $z=20$.

Now $m= y_{20}$, $i=21$, $y_{21} = y_{22} =m$, so, $p= x_{23}$ and $q= y_{23} = y_{24} = y_{25}$ define the initial cone, which is then restricted by $x_{24}$ and $p= x_{25}$ where it becomes a ray. For $\beta_{26}$, $p= x_{26} = x_{25}$ confirms the cone, but $y_{26}$ is ignored because the condition of Line 9 is satisfied and $y_{26}$ lies on the left outside the cone. Then $x_{27}$ and $y_{27}$ both lie on the left side outside the cone which violates the condition of Line 9, Line 18 yields $m_5= q= y_{25}$.

The next initial cone is given by $m= y_{25}$, $p= x_{27}$, $q= y_{27}$. Confirmation by $y_{28} = y_{29} = y_{30} = q$ and restrictions by $x_{29}$ and $p= x_{30}$ convert the cone into a ray. For $\beta_{31}$, $y_{31} = y_{30}$, $(m,y_{31},x_{31})$ is a left turn, violating the condition of Line 9, resulting in $m_6= q= y_{30}$.

Restarting with $m= y_{30}$, the initial cone is given by $p= x_{32}$ and $q= y_{32}= y_{33}$, it is restricted by $x_{33} = x_{34}$, $x_{35} = x_{36}$, $x_{37} = x_{38}$, $x_{39} = x_{40}= x_{41} = p$, confirmed by $y_{34} = y_{35}$, $y_{36} = y_{37}$, $y_{38} = y_{39}$ and $y_{40}$ and then restricted by $y_{41} = q$. Now $(m, y_{42}, x_{42})$ forms a left turn which violates the condition of Line 9, hence $m_7= p= x_{41}$.

For $m= x_{41} = x_{42}$, $p= x_{43} = x_{44} = x_{45}$ and $q= y_{43}$ determine the initial cone, which is restricted by $y_{44}$ and $q= y_{45}$ where the cone becomes a ray. Then $q= y_{46} = y_{45}$ confirms the cone but $x_{46}$ is ignored since it lies on the right outside the cone. Finally $x_{47}, y_{47}$ both lie on the right side outside the cone, resulting in $m_8= p= x_{45}$.

The next initial cone is found with $m= x_{45}$, $p= x_{46} = x_{47}$, $q= y_{46}$, it is confirmed by $x_{48} = x_{49}$ and $x_{50}$, restricted by $y_{47}= y_{48}$, $y_{49}= y_{50}= y_{51} =q$ and $p= x_{51}$. Then $(m, y_{52}, x_{52})$ forms a left turn, so, $m_9= q= y_{51}$ is found. 

For $m= y_{51} = y_{52}$, the initial cone is given by $p= x_{53}$ and $q= y_{53} = y_{54}$. It is confirmed by $y_{55} = y_{56} = y_{57} =q$ and constrained by $x_{54} = x_{55}$, $x_{56}$ and $x_{57} = p$ where the cone becomes a ray. Then $p= x_{57} = x_{58} = x_{59} = x_{60}$ confirms the cone, $y_{58}$ and $y_{59}$ are ignored (outside the cone to the left), $y_{60} = y_{61} = y_{62} = y_{63} =q$ confirms the cone, $x_{61}$ and $x_{62}$ are ignored (outside the cone to the right), $x_{63} =p$ confirms the cone. Finally $(m, y_{64}, x_{64})$ forms a left turn, the condition of Line 9 is not satisfied, so, $m_{10}= q= y_{63}$ is found. 

Algorithm \ref{alg:MPP-Alg-triangular} continues finding $m_{11}= y_{69}$, $m_{12}= y_{71}$ and $m_{13}= x_{77}$. The initial cone for $m= x_{77}$, given by $p= x_{78} = x_{79} = x_{80}$ and $q= y_{78}$, is restricted by $y_{79}$ and $y_{80}$ that converts the cone into a ray. Then $y_{80} = y_{81} = y_{82}$ and $y_{83} = q$ confirm the cone, $x_{81}$ and $x_{82} = x_{83}$ are ignored (outside the cone on the right), but $x_{84}, y_{84}$ both lie to the right outside the cone, consequently $m_{14}= p = x_{80}$. It is easy to verify that $m_{15}= x_{84}$.

Starting with $m= x_{84} = x_{85}$, $p= x_{86} = x_{87} = x_{88}$ and $q= y_{86}$ give the initial cone. It is constrained by $y_{87}$ and $y_{88} = y_{89} = y_{90} =q$ where it becomes a ray. Then $x_{89}$, $x_{90} = x_{91}$, $x_{92} = x_{93}$ and $x_{94}$ all are ignored (outside the cone on the right), $y_{91} = y_{92}$, $y_{93} = y_{94} = y_{95}$ and also $x{95}$ confirm the cone. Then Line 23 has the result $i=96$, which satisfies the Stop condition in Line 3, $m_{15}$ remains the last found polygon vertex. 

\rightline{$\vartriangle$}

As a result of Example 2, the polygon determined by Algorithm \ref{alg:MPP-Alg-triangular} coincides with the MPP that had been shown in Figure \ref{fig:Introd-example-regular-complex-MPP}, it is a weakly simple polygon that has thin degenerate parts. All points of type $x_i$ in the output list are concave MPP vertices, all points $y_i$ are convex MPP vertices. Several points are convex and concave vertices, for example, $y_{0}= x_{41}$ and $x_{10}= y_{30}$. The example shows that a convex MPP vertex is not necessarily a core vertex: the convex MPP vertices $y_0, y_{25}, y_{30}, y_{69}$ do not belong to the core of the complex.


\section{Correctness and complexity of the MPP Algorithm for regular complexes}
\label{sect:MPP-Alg-correctness}


Algorithm \ref{alg:MPP-Alg-triangular} adapts the earlier MPP algorithm proposed in \cite{Wiederhold2024} for rec\-tan\-gular tilings to the triangular tiling, but the correctness proofs require some distinct arguments. In a boun\-dary path of rectangular tiles, any triple of consecutive tiles either forms a straight move or corresponds to a convex or concave turn, a turn indicates the presence of an MPP vertex candidate \cite{Wiederhold2024}. In a boun\-dary path of triangular tiles, there is no relationship between these triples and MPP vertex candidates, no triple of consecutive tiles forms a straight move.

Due to Lemma \ref{lemma:regular-complex-hex}, a regular complex with at least two tiles actually has at least six tiles forming a hexagon, this justifies the suppositions of the following lemma and theorem. By Lemma \ref{lemma:regular-complex-starttile}, the leftmost tile $\beta_0$ of the top row of $\mathcal{C}$ is an upright triangle. This tile provides an initial MPP vertex:

\begin{lemma} 
For any regular complex $\mathcal{C}$ with at least six tiles in a triangular tiling, the lower right corner point of the upright triangle given as the leftmost tile in the top row of $\mathcal{C}$ is a convex vertex of the MPP of $\mathcal{C}$.
\label{lemma:MPP-initial-vertex}
\end{lemma}

\noindent Proof:
Let $T$ be the upright triangle that is the leftmost tile in the top row of $\mathcal{C}$, $T$ is a boundary tile. Denote by $m$ the lower right corner point of $T$ and by $P$ the MPP of $\mathcal{C}$.

\begin{figure}
\centering
\includegraphics[height=4.6cm]{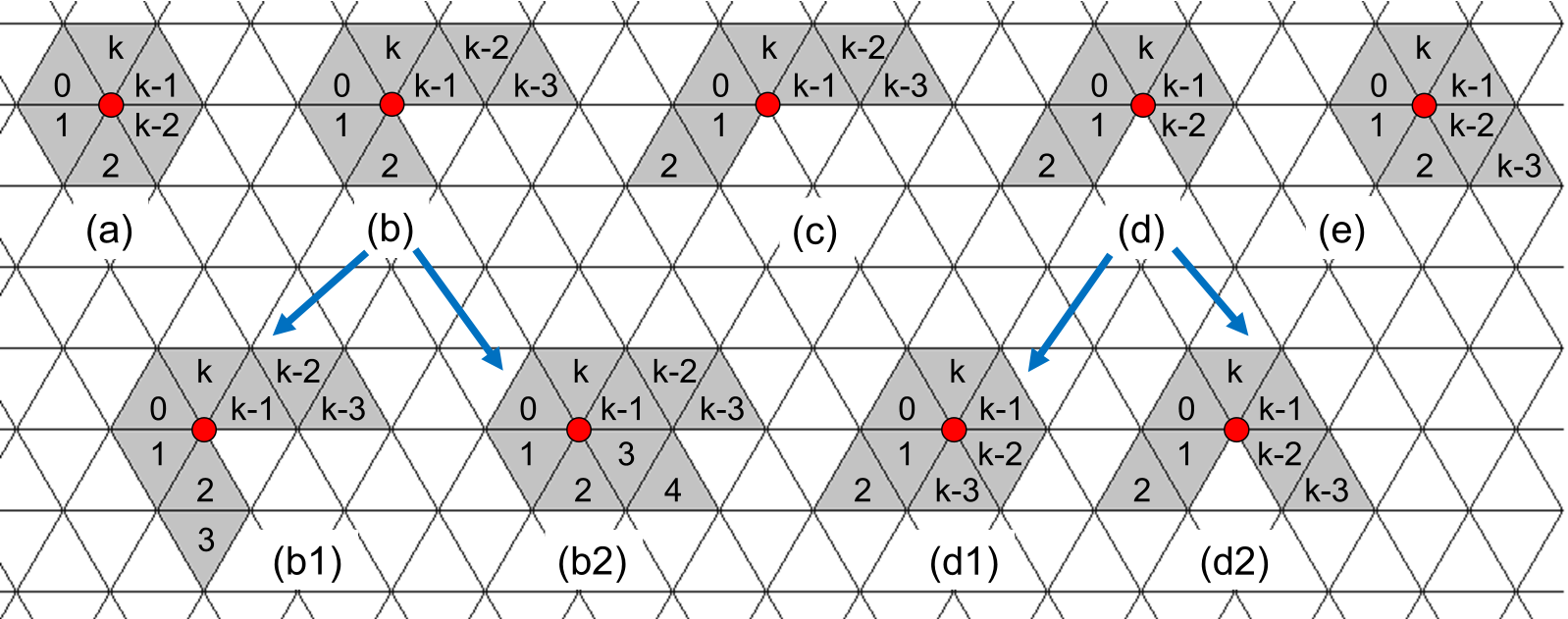}
\caption{All possible situations around the starting tile $T=\beta_0$ of the canonical boundary chain, discussed in the proof of Lemma \ref{lemma:MPP-initial-vertex}. Each $\beta_k$ is denoted by the number $k$, the right corner point $m$ of $T$ is drawn in red. (a) shows the case $\mathcal{C}= \mathcal{C}_{Hex}$ where $k=5$, in all other cases $\mathcal{C}$ has more than six tiles.}
\label{fig:MPP-initial-vertex}
\end{figure}

If $\mathcal{C}= \mathcal{C}_{Hex}$, see Figure \ref{fig:MPP-initial-vertex}(a), the claim is true because then $m$ is the unique vertex of $P$, which is a degenerate polygon consisting of the point $m$.

Now suppose that $\mathcal{C}$ has more than six tiles, then it has at least ten tiles and $P$ has at least two vertices. Let the cyclic list $\beta(\mathcal{C}) = (\beta_0 , \beta_1 , \cdots , \beta_k)$ be the canonical boundary path of $\mathcal{C}$ such that $\beta_0 =T$.

By Lemma \ref{lemma:canonic-boundary-leaves-C-onLeft}, $\beta(\mathcal{C})$ is regular, in particular, $\beta_1 \neq \beta_k$, and both tiles are adjacent to $\beta_0$. Since $\beta(\mathcal{C})$ traces the boundary counterclockwise and the left neighbor of $\beta_0$ does not belong to $\mathcal{C}$, $\beta_1$ is the neighbor below $\beta_0$, and $\beta_k$ is the right neighbor of $\beta_0$. By the regularity of $\beta(\mathcal{C})$, $\beta_{k-1} \neq \beta_0$, which forces $\beta_{k-1}$ to be the right neighbor of $\beta_k$, see Figure \ref{fig:MPP-initial-vertex}. Considering that $\mathcal{C}\neq \mathcal{C}_{Hex}$, $\beta_2$ can be the left or right neighbor of $\beta_1$, and $\beta_{k-2}$ can be the neighbor below or to the right of $\beta_{k-1}$, the latter forces $\beta_{k-3}$ to be the right neighbor of $\beta_{k-2}$. This produces the situations (b),(c),(d),(e) of Figure \ref{fig:MPP-initial-vertex}. In (b), $\beta_3$ can be the neighbor below $\beta_2$, yielding the case (b1), or the right neighbor of $\beta_2$, which forces $\beta_4$ to lie to the right from $\beta_3$, this is case (b2). In (c),(d),(e), $\beta_3$ may be the neighbor lying below or on the left of $\beta_2$. Cases (d1) and (d2) are generated from case (d) where $\beta_{k-3}$ can be the left or right neighbor of $\beta_{k-2}$.

Let $\gamma$ denote the frontier curve of $P$. By Lemma \ref{lemma:MPP-shortest-curve-through-canon-bound-path}, $\gamma$ is the shortest polygonal curve that passes through $\beta(\mathcal{C})$. In particular, it visits the tiles $\beta_{k-3}$, $\beta_{k-2}$, $\beta_{k-1}$, $\beta_k$, $\beta_0$, $\beta_1$, $\beta_2$, $\beta_3$, in that order. Let $h$ be the horizontal straight line that contains $m$. Figure \ref{fig:MPP-initial-vertex} shows all possible situations around the point $m$. Since $\beta_0$ is the leftmost tile of the top row of $\mathcal{C}$, the curve $\gamma$ has neither points above the line $h$, nor to the left of $m$ on the line $h$. Hence $\gamma$ visits $\beta_{k-1}$, $\beta_k$, $\beta_0$, $\beta_1$, all via the point $m$, as this is the shortest way, and then continues via $\beta_2$, $\beta_3$, $\cdots $, where $\gamma$ lies strictly below the line $h$ or passes on $h$ returning to $m$ from the right. In particular, $\gamma$ cannot have a horizontal segment on $h$ on the left side of $m$. As a consequence, $\gamma$ has $m$ as a vertex, which is evidently convex, that is, $m$ is a convex vertex of $P$.

\qed
\medskip

The lower right corner $m$ of the leftmost triangle tile of the top row of $\mathcal{C}$, which provides a special convex MPP vertex by Lemma \ref{lemma:MPP-initial-vertex}, is not necessarily a vertex of the core of $\mathcal{C}$. For example, in the complex of Figure \ref{fig:Introd-example-regular-complex-MPP}, $m=d$ is a convex (and also concave) MPP vertex but does not belong to the core.

\begin{theorem} 
For any regular complex $\mathcal{C}$ with at least six tiles in a triangular tiling, using the canonical boundary path of $\mathcal{C}$ as input list, Algorithm \ref{alg:MPP-Alg-triangular} computes the ordered sequence of vertices of the MPP of $\mathcal{C}$, due to counterclockwise tracing of the MPP frontier curve.
\label{theorem-MPP}
\end{theorem}

\noindent Proof:
Let $\beta (\mathcal{C})  = (\beta_0 ,\beta_1 ,\cdots ,\beta_t )$ be the canonical boundary path of $\mathcal{C}$.

\smallskip\noindent (1) The smallest possible case is the complex $\mathcal{C}_{Hex}$ shown in Figure \ref{fig:example1}(a). Its core is a single point being the common vertex of the six tiles. It coincides with the MPP correctly computed by Algorithm \ref{alg:MPP-Alg-triangular}, as shown in Example 1. Now assume that $\mathcal{C}$ has at least seven tiles. 

\smallskip\noindent (2) By Lemma \ref{lemma:regular-complex-hex}, $\mathcal{C}$ is a union of hexagons $\mathcal{C}_{Hex}$ and, eventually, simple paths connecting parts that are unions of such hexagons. When tracing the canonical boundary path $\beta (\mathcal{C})$, the endpoint $y_k$ (or $y_i$) used in Algorithm \ref{alg:MPP-Alg-triangular} may lie on the frontier of \textit{core}$(\mathcal{C} )$, but can also belong to $fr(|\mathcal{C} |)$ when passing through a narrow part of $\mathcal{C}$ that does not contain any hexagon $\mathcal{C}_{Hex}$. Nevertheless, due to Lemma \ref{lemma:canonic-boundary-leaves-C-onLeft}, the endpoint $x_k$ (or $x_i$) always lies on $fr(|\mathcal{C} |)$.

\smallskip\noindent (3) When Algorithm \ref{alg:MPP-Alg-triangular} starts with the last found polygon vertex $m_n$ provided by some $\beta_z$, and, after constructing and successively restricting the cone, finds the next vertex $m_{n+1}$ as the point $p$ or $q$ given by some tile $\beta_j$, $j\geq z+1$, the cone construction guaran\-tees that $\overline{m_n m_{n+1}}\subset\beta_{z}\cup \beta_{z+1}\cup \cdots \cup \beta_j$. Hence the constructed polygonal curve passes through all boundary tiles in the order specified in $\beta (\mathcal{C})$ and does not leave the boundary. The remaining parts of the proof will show that the list of vertices determined by Algorithm \ref{alg:MPP-Alg-triangular} coincides with the ordered sequence of all MPP vertices. In particular, this means then that the polygonal curve obtained from the algorithm is weakly simple.

\smallskip\noindent (4) By Lemma \ref{lemma:regular-complex-starttile}, the leftmost tile $\beta_0$ of the top row of $\mathcal{C}$ is an upright triangle. Due to Lemma \ref{lemma:MPP-initial-vertex}, its lower right corner $m=m_1$ is an MPP vertex.

\smallskip\noindent (5) Now let $m_n$, $n\geq 1$, be any polygon vertex determined by Algorithm \ref{alg:MPP-Alg-triangular}, $m_n$ is found as a point $x_z$ or as a point $y_z$. Suppose that $m_n$ is an MPP vertex. Denote by $a$ the next MPP vertex after $m_n$, for counterclockwise tracing of the MPP frontier curve. In the following we will prove that $a$ coincides with the next polygon vertex $m_{n+1}$ found by Algorithm \ref{alg:MPP-Alg-triangular}. 

\begin{figure}
\centering
\includegraphics[height=3.5cm]{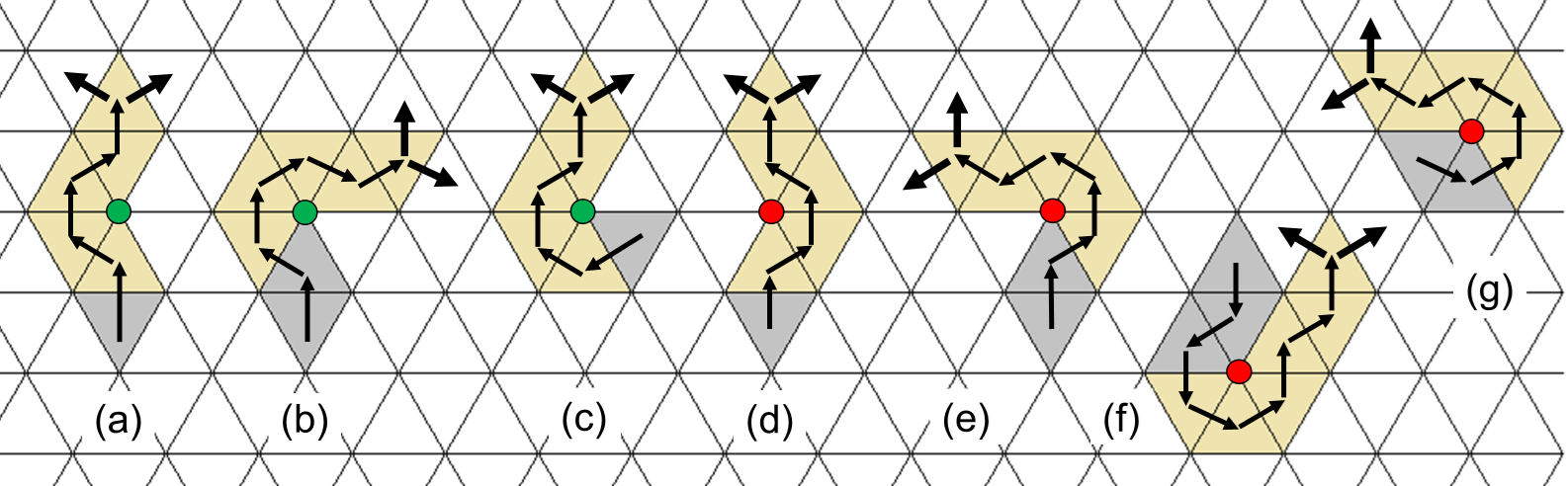}
\caption{All possible situations around a concave MPP vertex shown as a green point in (a-c), or a convex MPP vertex shown as a red point in (d-g), discussed in the proof of Theorem \ref{theorem-MPP}. Grey and beige filled tiles belong to the canonical boundary path, the tracing of which is indicated by arrows. At the end, bold arrows illustrate the two continuation possibilities. In (a-c), the green point $m_n= x_z$ is a concave vertex of $fr(|\mathcal{C} |)$. In (d-e), the red point $m_n= y_z$ can be a convex vertex of \textit{core}$(|\mathcal{C}|)$ or a concave vertex of $fr(|\mathcal{C}|)$ (if $\beta (\mathcal{C})$ passes through a thin part of the complex), but it is a convex vertex of \textit{core}$(|\mathcal{C}|)$ in (f-g). Each figure shows one rotational instance of a situation, another instance resulting from a rotation by a multiple of $60^{\circ }$ could present a path starting with a type of triangle (upright or inverted) distinct from that shown in the figure.}
\label{fig:preliminar-for-initial-cones}
\end{figure}

\begin{figure}
\centering
\includegraphics[height=3.5cm]{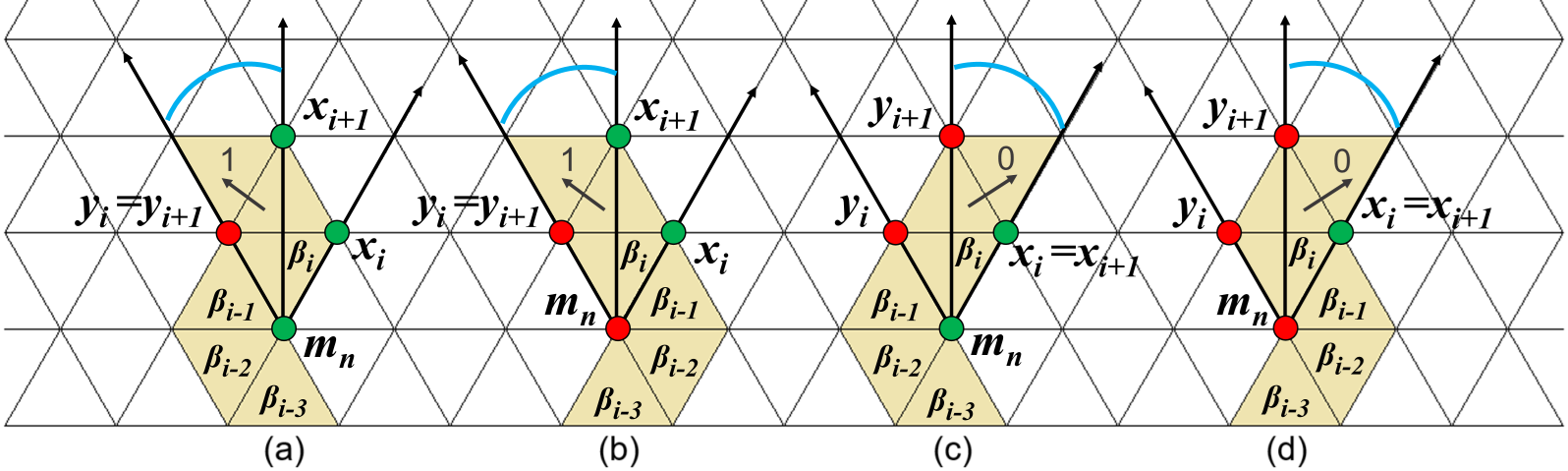}
\caption{To illustrate the proof of Theorem \ref{theorem-MPP}, boundary path tiles are filled beige. The cone initialized by $x_i$, $y_i$ is always of $60^{\circ}$ but is restricted by $x_{i+1}$ or $y_{i+1}$ to a cone of $30^{\circ}$, as indicated by the blue arcs. The polygon vertex $m_n$ is concave in (a) and (c), convex in (b) and (d).}
\label{fig:initial-cones}
\end{figure}

\smallskip\noindent (5a) If $m_n= x_z$, $m_n$ is a concave vertex of the MPP and of $fr(|\mathcal{C}|)$. The five beige filled boundary tiles in each of the Figures \ref{fig:preliminar-for-initial-cones}(a-c) presents the same situation except for a rotation by a multiple of $60^{\circ}$ around $m_n$, this part is copied to Figures \ref{fig:initial-cones}(a)(c). If $m_n= y_z$ then $m_n$ is a convex MPP vertex, see each of the Figures \ref{fig:preliminar-for-initial-cones}(d-g) where the beige filled tiles present the same situation except for a rotation by a multiple of $60^{\circ}$ around $m_n$, these five tiles are copied to Figures \ref{fig:initial-cones}(b)(d). In each part of Figure \ref{fig:initial-cones}, a sixth tile is added showing the possible continuations.

Algorithm \ref{alg:MPP-Alg-triangular} initializes a cone of visibility through upcoming boundary tiles, rooted at $m_n$ with right border $\overrightarrow{m_np}$ and left border $\overrightarrow{m_nq}$, where $p=x_i\in fr(|\mathcal{C}|)$ and $q=y_i$ for some $i\geq z+1$. If $i>z+1$, $\beta_{z+1} ,\beta_{z+2} ,\cdots , \beta_i$ all have $m_n$ as vertex and surround $m_n$ counterclockwise if $m_n$ is convex and clockwise if $m_n$ is concave. Figure \ref{fig:initial-cones} shows the two possible situations for concave $m_n$ in (a)(c), and the two possibilities for $m_n$ convex in (b)(d). Any other situation can be obtained via a rotation around $m_n$ by a multiple of $60^{\circ}$; after such a rotation, $\beta_{i-3}$ may be an upright or an inverted triangle.

\smallskip\noindent (5b) As a result of (5a) of this proof, the initial cone is one of the sextants rooted at $m_n$. We will only discuss the case of the sextant which opens upwards, as shown in all parts of Figure \ref{fig:initial-cones}, movement directions and relative positions between tiles we will use in all the following, are valid only for this sextant. The arguments would be similar for the other sextants. Since $\beta (\mathcal{C})$ is regular, from the upright triangle $\beta_{i+1}$ it continues to the right or to the left, Figure \ref{fig:initial-cones} shows both options, the cone is constrained to $30^{\circ}$ by $x_{i+1}$ or $y_{i+1}$. The cone is eventually further restricted until, for some $k>i$, $\beta_k$ is the first tile that does not satisfy the condition of Line 9 of Algorithm \ref{alg:MPP-Alg-triangular}, i.e., all tiles $\beta_i, \cdots , \beta_{k-1}$ satisfy the condition of Line 9. By construction, the cone then passes through all tiles $\beta_i ,\beta_{i+1} ,\cdots , \beta_{k-1}$, and $m_{n+1}$ is provided by $\beta_j$ with $i\leq j<k$. 

\begin{figure}
\centering
\includegraphics[height=3.2cm]{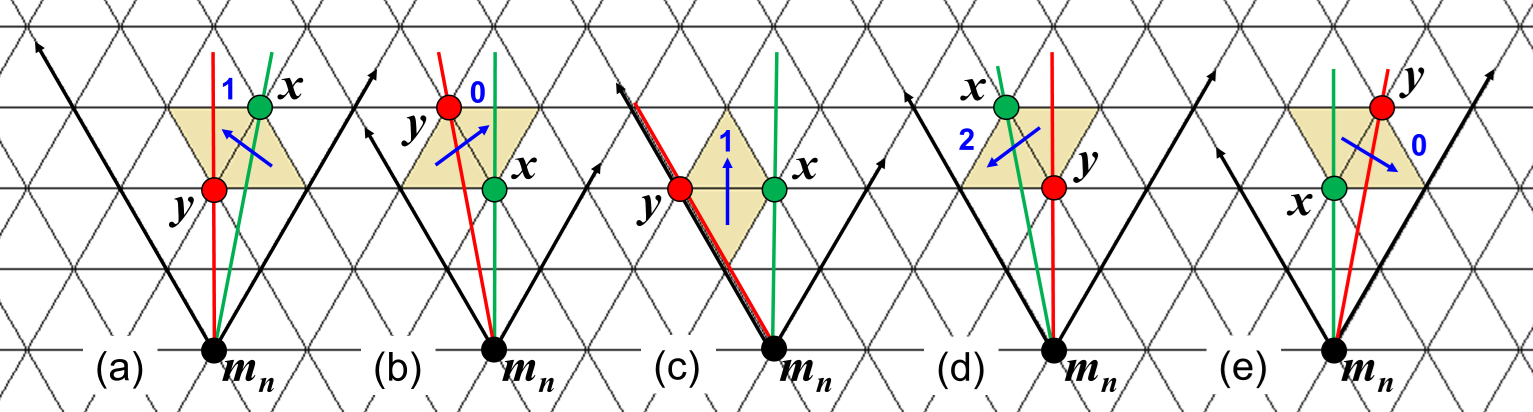}
\caption{For (5c) of the proof of Theorem \ref{theorem-MPP}, consider the cone rooted at $m_n$ with the borders drawn by black arrows and a boundary path containing the two beige filled tiles where blue arrows and numbers indicate the Freeman path direction from $\beta_s$ to $\beta_{s+1}$. If $x$ is the right endpoint of $\beta_s\cap \beta_{s+1}$ and $y$ its left endpoint, the triple $(m_n, y, x)$ forms a right turn in (a), (b) and (c), but it is a left turn in (d) and (e) if both $\beta_s , \beta_{s+1}$ lie inside the cone. In (d) and (e), if only $\beta_s$ lies inside the cone, $x,y$ both belong to the left cone border in (d), and to the right border in (e), then $m_n, y, x$ are collinear.}
\label{fig:cone-richtig-falsch}
\end{figure}

\smallskip\noindent (5c) A cone resulting from restrictions of the sextant which opens upwards, is still rooted at $m_n$ and opens upwards; actually it lies within a cone with an angle of $30^{\circ}$, as shown in (all parts of) Figure \ref{fig:initial-cones}. Within such a cone, for an inverted triangle $\beta_s\in \beta (\mathcal{C})$, if $\beta_{s+1}\in \beta (\mathcal{C})$ is above $\beta_s$ (in Freeman direction $1$) then $(m_n,y_s,x_s)$ forms a right turn, see Figure \ref{fig:cone-richtig-falsch}(c). But if $\beta_{s+1}$ lies in Freeman direction $0$ or $2$ and belongs to the cone then $(m_n,y_s,x_s)$ is a left turn, which violates the condition of Line 9, see Figure \ref{fig:cone-richtig-falsch}(d)(e). For an upright triangle $\beta_s\in \beta (\mathcal{C})$, see Figure \ref{fig:cone-richtig-falsch}(a)(b), if $\beta (\mathcal{C})$ moves to $\beta_{s+1}$ in Freeman direction $0$ or $1$ then $(m_n,y_s,x_s)$ is a right turn. Algorithm \ref{alg:MPP-Alg-triangular} could not find a downward movement of $\beta (\mathcal{C})$ since this would require a previous movement from an inverted triangle $\beta_{s-1}$ to $\beta_s$ in Freeman direction $0$ or $2$, but then $\beta_{s-1}$ would have already violated the condition of Line 9.

\begin{figure}
\centering
\includegraphics[height=5.6cm]{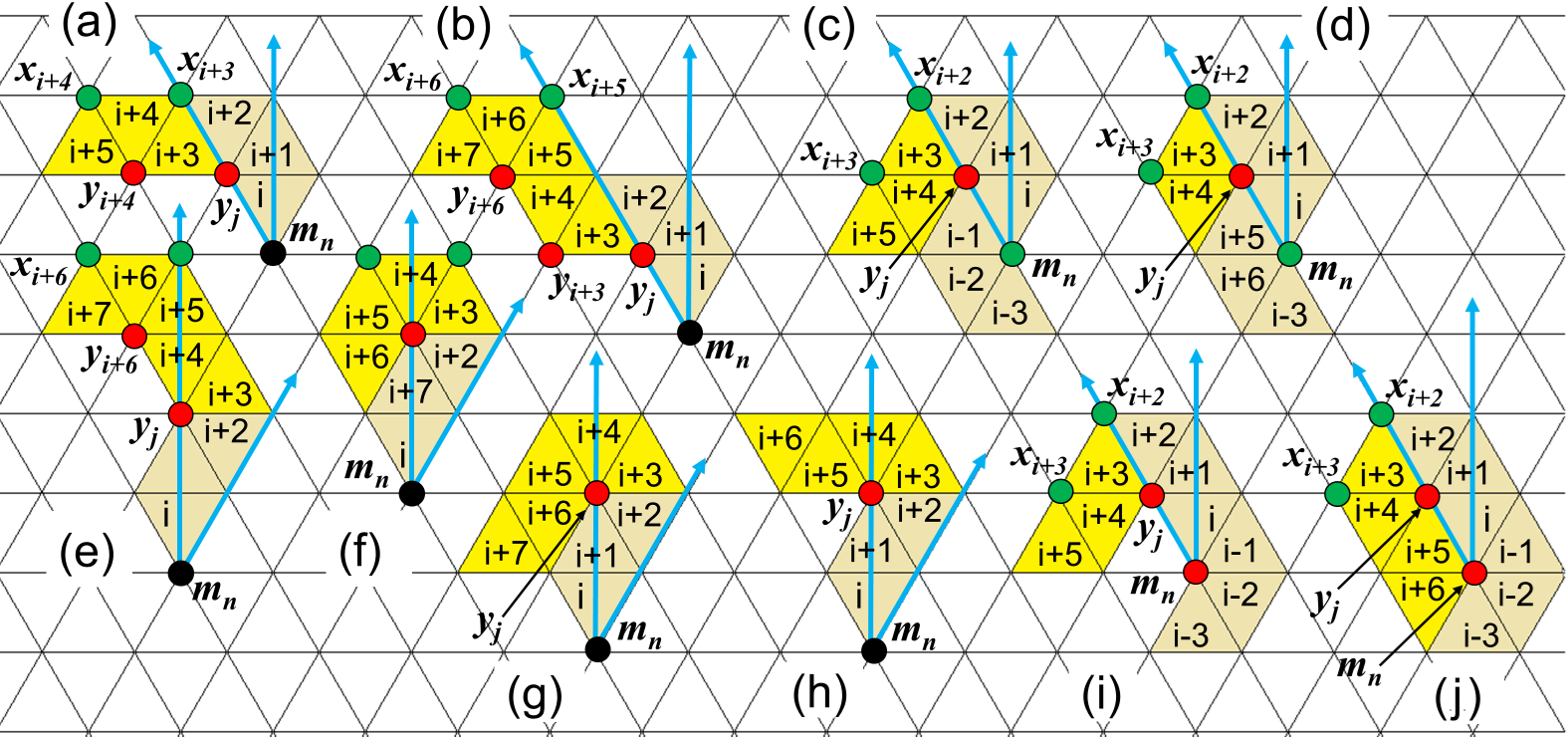}
\caption{Smallest situations for an essential boundary movement to the left, detected at $\beta_k$, for (5d) of the proof of Theorem \ref{theorem-MPP} where $\mathcal{C}$ is assumed to have more than six tiles. A number like $i+1$ indicates the tile $\beta_{i+1}$. Beige filled tiles are copied from initial situations shown in Figure \ref{fig:initial-cones}, these and the yellow filled tiles belong to the boundary path $\beta (\mathcal{C} )$. Parts (a) and (b) correspond to situations from Figures \ref{fig:initial-cones}(a)(b), (e-h) deal with the cases from Figures \ref{fig:initial-cones}(c)(d). See the discussion in the text.}
\label{fig:links-kleinste}
\end{figure}

\smallskip\noindent (5d) Figure \ref{fig:links-kleinste} shows the smallest situations where an essential boundary movement to the left is detected at $\beta_k$, starting with all possible cones initialized by $m_n, x_i, y_i$ from Figure \ref{fig:initial-cones}. Recall that $\beta (\mathcal{C})$ is regular, i.e., $\beta_{i-1} \neq \beta_{i+1}$ for every $i$, and $\beta (\mathcal{C})$ does not cross itself. Let us discuss the cases of Figure \ref{fig:links-kleinste}: 

\noindent Fig.(a): $y_i = y_{i+1} = y_{i+2} = q$ is the last point that confirms the left cone border, with $x_{i+2}$ the cone becomes a ray which is confirmed by $p= x_{i+3}$, $y_{i+3}$ is ignored. But $x_{i+4}, y_{i+4}$ both lie outside the cone on the left, hence $m_{n+1} = q = y_{i+2}$, $j=i+2$, $k=i+3$. There is no straight line segment from $m_n$ inside the boundary, i.e., passing through $\beta_{i}, \beta_{i+1}, \beta_{i+2}, \beta_{i+3}, \beta_{i+4}$, which reaches the tile $\beta_{i+5}$.

\noindent Fig.(b): the situation is equal to Fig.(a) until $\beta_{i+4}$, $x_{i+4}$ and $x_{i+5}$ confirm the right border, and $y_{i+4}= y_{i+5}$ is ignored. But $x_{i+6}, y_{i+6}$ both lie on the left outside the cone, so $m_{n+1} = q = y_{i+2}$, $j=i+2$, $k=i+6$. There is no straight line from $m_n$ inside the boundary that reaches $\beta_{i+7}$.

\noindent Fig.(c): $y_i = y_{i+1} = y_{i+2} = q$ defines and confirms the left cone border,  $p= x_{i+2}$ restricts the cone to become a ray. Then $(m_n, y_{i+3}, x_{i+3})$ is the left turn (violating the condition of Line 9 of Algorithm \ref{alg:MPP-Alg-triangular}), so $m_{n+1} = q = y_{i+2}$, $j=i+2$, $k=i+3$. It is clear from the figure that no straight line from $m_n$ inside the boundary can reach $\beta_{i+5}$.

\noindent Fig.(d): it is equal to Fig.(c) up to $\beta_{i+4}$, $\beta_{i+5} = \beta_{i-1}$ is the unique possible distinct continuation. Since $\beta_{i+6} \neq \beta_{i+4}$ and $\beta_{i+6} \neq \beta_{i}$ ($\beta (\mathcal{C})$ does not cross itself), $\beta_{i+6} = \beta_{i-2}$. The shortest curve travelling through $\beta_{i}, \beta_{i+1}, \beta_{i+2}, \cdots $ until reaching $\beta_{i+6}$ needs $y_j$ as convex vertex, since no straight line from $m_n$ passing through these tiles reaches $\beta_{i+6}$.

\noindent Fig.(e): $y_{i+1} = y_{i+2} = y_{i+3} = q$ and $x_{i+3} = x_{i+4}, x_{i+5} =p$ restrict the cone, it becomes a ray, $y_{i+4} = y_{i+5}$ is ignored. Then $(m_n, y_{i+6}, x_{i+6})$ forms the left turn, hence $m_{n+1} = q = y_{i+3}$, $j=i+3$, $k=i+6$. Continuing $\beta (\mathcal{C})$ as in the figure, there is no straight line from $m_n$ inside the boundary to reach $\beta_{i+7}$.

\noindent Fig.(f): $y_{i+1} = y_{i+2} = y_{i+3} = q$ and $p= x_{i+3}$ restrict the cone, but $(m_n, y_{i+4}, x_{i+4})$ is a left turn, so $m_{n+1} = q = y_{i+3}$, $j=i+3$, $k=i+4$. If $\beta (\mathcal{C})$ continues until $\beta_{i+7} = \beta_{i+1}$ as indicated, $\beta_{i+8}$ must be equal to $\beta_{i}$, then no straight line lying inside $\beta (\mathcal{C})$ and starting at $m_n$ reaches $\beta_{i+8}$.

\noindent Fig.(g): the situation is equal to Fig.(f) until $\beta_{i+6}$, $j=i+3$, $k=i+4$. Now, $\beta_{i+7}$ cannot be reached by a straight line that starts at $m_n$ and follows $\beta (\mathcal{C})$.

\noindent Fig.(h): it is equal to Fig.(g) up to $\beta_{i+6}$, $j=i+3$, $k=i+4$, but with $\beta_{i+6}$ as indicated, this tile cannot be reached by a straight line lying within $\beta (\mathcal{C})$ and starting at $m_n$.

\noindent Fig.(i): $q= y_{i+2}$ confirms the cone, $p=x_{i+2}$ converts it into a ray, $m_n, y_{i+2}, x_{i+2}$ are collinear. But $(m_n, y_{i+3}, x_{i+3})$ is the left turn, hence  $m_{n+1} = q = y_{i+2}$, $j=i+2$, $k=i+3$. If $\beta (\mathcal{C})$ continues as shown in the figure, there is no straight line lying inside $\beta (\mathcal{C})$ that starts at $m_n$ and reaches $\beta_{i+5}$.

\noindent Fig.(j): the situation is equal to Fig.(i) until $\beta_{i+4}$, $j=i+3$, $k=i+4$. The position of $\beta_{i+5}$ forces $\beta_{i+6}$ to lie below, then no straight line starting at $m_n$ and following $\beta (\mathcal{C})$ reaches $\beta_{i+6}$.

\smallskip
As result of analysing the smallest cases of Figure \ref{fig:links-kleinste}, first we learn that $j\geq i+2$. Moreover, the last point $q=y_j$ that restricts or confirms the left cone border, provides the polygon vertex $m_{n+1}$, and there is some $v>j$ such that no straight line starting at $m_n$ and passing through $\beta_{i}, \beta_{i+1}, \beta_{i+2}, \cdots , \beta_{i+v-1}$ (in this order) reaches $\beta_{i+v}$. This implies that the MPP frontier, being the shortest polygonal curve that follows the boundary, needs $y_j$ as a convex vertex. But $\overrightarrow{m_ny_j}$ lies within the boundary, hence $a=m_{n+1} = y_j$.

\begin{figure}
\centering
\includegraphics[height=5.7cm]{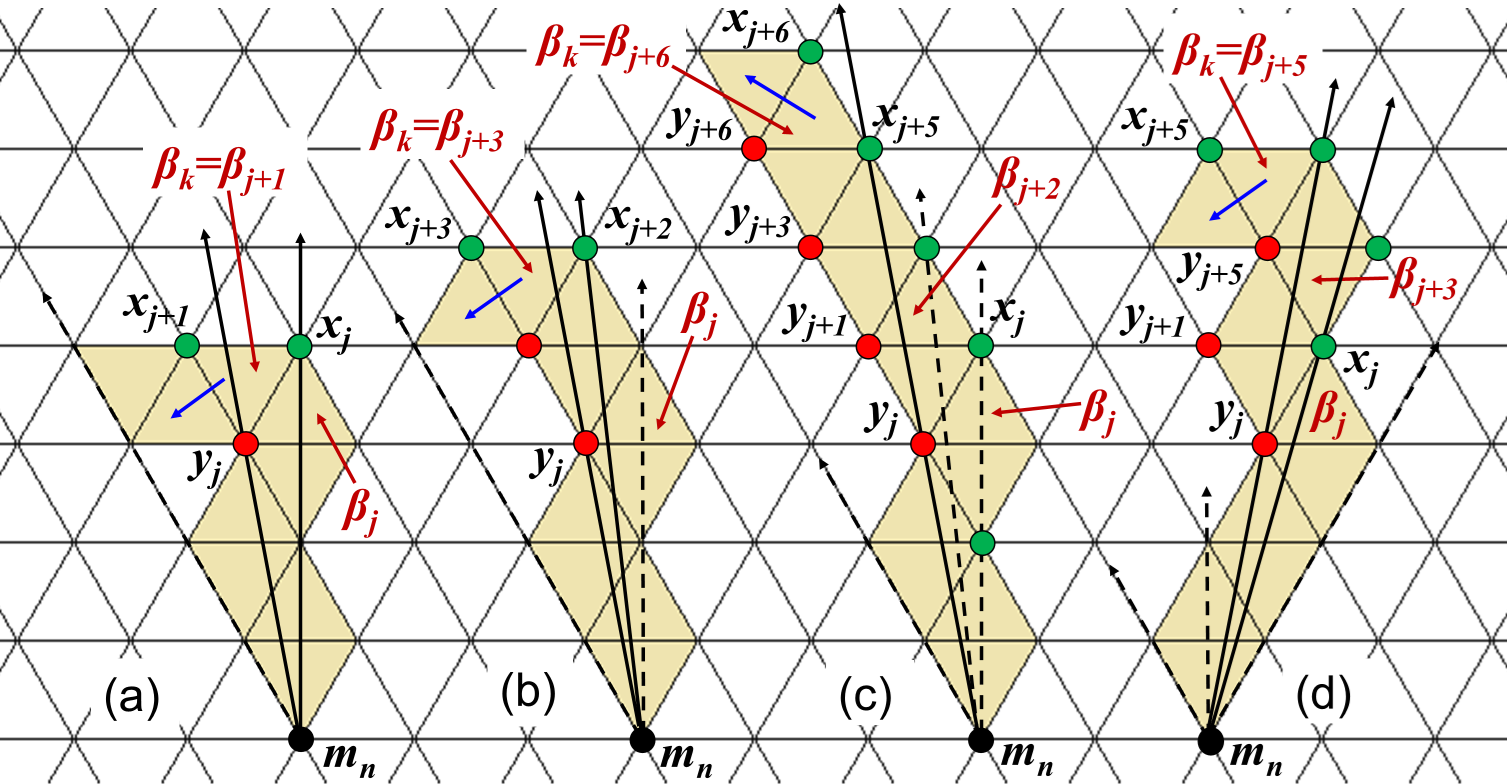}
\caption{Illustration for (5e) of the proof of Theorem \ref{theorem-MPP}. Algorithm \ref{alg:MPP-Alg-triangular} detects an essential boundary movement to the left at $\beta_k$ when the left cone border is determined by $q=y_j$, $\beta_k$ is the first tile that violates the condition of Line 9. This can be caused by $x_k$, $y_k$ both being outside the cone to the left as in (b),(c),(d), or when $(m_n, y_k, x_k)$ forms a left turn as in (a),(b),(d). In general, $j+1\leq k$, (a) shows a case for $k=j+1$. In (b) $k=j+3$, where $y_{j+1} = y_{j+2}$ lies outside the cone to the left and hence is ignored. In (c) $k=j+6$, $x_j$, $x_{j+2}$ and $x_{j+5}$ constrain the cone to become a ray, but $y_{j+1}, y_{j+2},\cdots , y_{j+5}$ are all ignored because they lie outside the cone to the left. Similarly, $k=j+5$ in (d) where $y_{j+1}$ and  $y_{j+2} = y_{j+3} = y_{j+4}$ are ignored.}
\label{fig:nach-links}
\end{figure}

\smallskip\noindent (5e) If an essential boundary movement to the left is detected at $\beta_k$, this tile violates the condition of Line 9, which means that both $x_k$ and $y_k$ lie outside the cone on the left, or $(m_n,y_k,x_k)$ forms a left turn and $x_k$ lies on the left of $\overrightarrow{m_np}$. Moreover, $m_{n+1} = y_j$ is a convex vertex of the polygon determined by Algorithm \ref{alg:MPP-Alg-triangular} and the last point that restricts or confirms the left cone border among all endpoints $y_r$ of edges $e_r=\beta_r\cap\beta_{r+1}$ for $r\in \{ i+1, \cdots , k-1\}$, that is, all points $y_{j+1}, y_{j+2},\cdots ,y_{k-1}$ lie strictly on the left of the left border $\overrightarrow{m_ny_j}$, see Figure \ref{fig:nach-links}. The arguments used in what follows to show that $a= m_{n+1} = y_j$, are similar to the discussion of the smallest situations in (5d).

\smallskip\noindent (5e-1) Only two kinds of moves of $\beta (\mathcal{C})$ can lead to both $x_k$ and $y_k$ being outside the cone on the left:
if $\beta_k$ is an upright triangle and $\beta (\mathcal{C})$ moves up-left from $\beta_k$ to $\beta_{k+1}$ as in Figure \ref{fig:nach-links}(c), or, if $\beta_k$ is an inverted triangle and $\beta (\mathcal{C})$ moves down-left from $\beta_k$ to $\beta_{k+1}$ as in Figure \ref{fig:nach-links}(b)(d). In both cases, the whole tile $\beta_{k+1}$ lies strictly outside the cone on the left. Hence there is no straight line that starts at $m_n$ and passes through all tiles $\beta_i , \cdots , \beta_j ,\cdots , \beta_k$ until reaching $\beta_{k+1}$. Hence the MPP frontier, which is the shortest curve passing from $m_n$ through all these tiles until $\beta_{k+1}$, requires $y_j$ as a convex vertex after $m_n$. Since the segment $\overline{m_n y_j}$, which is the shortest path between these two points, lies inside the boundary, it follows that $a= m_{n+1} = y_j$.

\smallskip\noindent (5e-2) Using the results of (5c) of this proof, if $(m,y_k,x_k)$ is found as forming a left turn (and $x_k$ lies on the left of $\overrightarrow{m_np}$), then $\beta_k$ is an inverted triangle inside the current cone from where $\beta (\mathcal{C})$ moves downward-left to $\beta_{k+1}$, which lies inside the initial cone, as in Figure \ref{fig:cone-richtig-falsch}(d). The case that both $x_k, y_k$ lie outside the cone on the left has already been discussed above. Assume now $x_k$ lying strictly on the left of $\overrightarrow{m_ny_k}$, and that $y_k$ is not strictly outside the current cone on the left. 

This implies $k=j+1$. To see this, suppose to the contrary that $k\geq j+2$. Since for all $r\in \{ i, i+1, \cdots , k-1\}$, $\beta_r$ satisfies the condition of Line 9, $\beta (\mathcal{C})$ moves from $\beta_r$ to $\beta_{r+1}$ in upward direction as in Figure \ref{fig:cone-richtig-falsch}(a-c). To find that $\beta_{k+1}$ is the downward-left neighbor of $\beta_k$, $\beta (\mathcal{C})$ must perform a movement upward-left from the upright triangle $\beta_{k-1}$ to $\beta_k$, and hence an upward movement from the inverted triangle $\beta_{k-2}$ to $\beta_{k-1}$. Then $y_k= y_{k-1} = y_{k-2}$, which contradicts our assumption that $y_{j+1}, y_{j+2}, \cdots , y_{k-1}$ all lie on the left of $\overrightarrow{m_n y_j}$, in particular, $y_{k-1}$ lies strictly on the left of the left cone border. This proves $k=j+1$.

The case $k=j+1$ presents a situation similar to Figures \ref{fig:links-kleinste}(c)(d)(f-j) and \ref{fig:nach-links}(a). No matter how $\beta (\mathcal{C})$ continues, clearly there exists $v>j$ such that no straight line that starts at $m_n$ and passes through $\beta_{i}, \beta_{i+1}, \beta_{i+2}, \cdots , \beta_{i+v-1}$ (in this order) can reach $\beta_{i+v}$. Therefore, the MPP frontier curve needs $y_j$ as a convex vertex, using that $\overrightarrow{m_ny_j}$ lies in the boundary, it follows that $a= m_{n+1} = y_j$.

\smallskip\noindent (5f) If an essential boundary movement to the right is detected at $\beta_k$, $m_{n+1}$ is a concave vertex of $fr(|\mathcal{C} |)$, see Figures \ref{fig:preliminar-for-initial-cones}(a-c) and \ref{fig:initial-cones}(a)(c). Now $x_j$ is the last point that constrains or confirms the right cone border, among all endpoints $x_r$ of edges $e_r=\beta_r\cap\beta_{r+1}$, $r\in \{ i+1, \cdots , k-1\}$, that is, $x_{j+1}, x_{j+2},\cdots ,x_{k-1}$ all lie strictly on the right of the current right cone border $\overrightarrow{m_nx_j}$. Arguments analogous to those used in (5d) and (5e) lead to the result that $a$, which is the next MPP vertex after $m_n$, coincides with $m_{n+1} = x_j$.

\smallskip\noindent (6) It is easy to see that Algorithm \ref{alg:MPP-Alg-triangular} terminates correctly when the input list has been processed: if $\beta_0$ was treated again after finding the last polygon vertex, Algorithm \ref{alg:MPP-Alg-triangular} would again find the first vertex $m_1$. Hence Algorithm \ref{alg:MPP-Alg-triangular} correctly computes the ordered list of all vertices of the MPP of $\mathcal{C}$.

\qed
\medskip

To analyse the time complexity of Algorithm \ref{alg:MPP-Alg-triangular}, note that it takes as input the ca\-no\-ni\-cal boundary path $\beta (\mathcal{C})$ that can be obtained by Algorithm \ref{alg:bound-tracing}, which has linear time complexity in terms of the number of tiles of $\beta (\mathcal{C})$. Both algorithms do not require to examine the entire complex $\mathcal{C}$, except in the case where $\mathcal{C}$ coincides with its boundary. 

\begin{figure}
\centering
\includegraphics[height=5.5cm]{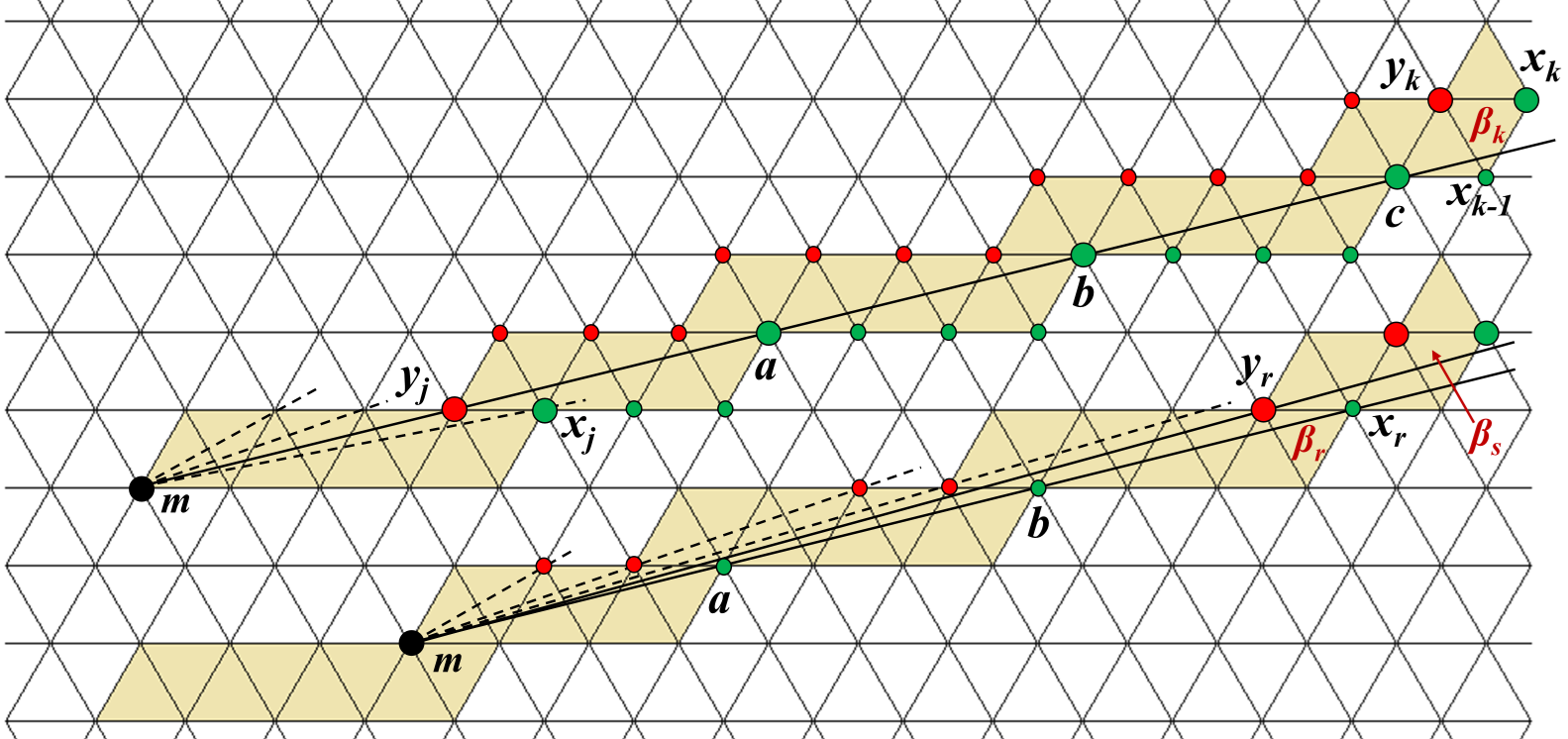}
\caption{In the figure above, the small red points $y_i$ are ignored by Algorithm \ref{alg:MPP-Alg-triangular} because they lie outside the cone to the left. Likewise, the small green points $x_i$ are outside the cone to the right and are therefore ignored. Starting from the last MPP vertex $m$ found, detecting the next MPP vertex $y_j$ provided by $\beta_j$ requires examining $\beta_{j+1}, \beta_{j+2}, \cdots , \beta_k$ until $\beta_k$ no longer satisfies the condition of Line 9, where $k-j$ is a large number. The right cone border is constrained by $x_j$ and $a$ and then confirmed by $b$ and $c$, while $y_j$ continues to determine the left cone border. Finally, $x_k, y_k$ are both outside the cone on the left, so $m=y_j$ is updated. In the figure below, restarting at $m$, the left cone border is restricted by several points shown as small red disks until $y_r$ determines this border. The right border is constrained by $a$ and confirmed by $b$ and $x_r = x_{r+1} = x_{r+2}$. For $s=r+3$, $x_s, y_s$ lie both outside the cone on the left, which allows finding the next MPP vertex $y_r$, there is a small backtracking by $s-r=3$ tiles.}
\label{fig:complexity}
\end{figure}

While processing the input list $\beta (\mathcal{C})$, Algorithm \ref{alg:MPP-Alg-triangular} traces back on this list to the last point $p=x_j$ that determines the right cone border $\overrightarrow{m_np}$, or to $q=y_j$ that specifies the left cone border $\overrightarrow{m_nq}$, each time an essential boundary movement is detected. Consider the input list $\beta (\mathcal{C}) = (\beta_1, \beta_2 ,\cdots , \beta_s)$ of boundary tiles such that $\beta_1$ provides the first MPP vertex $m_1$. Suppose that there is a worst case where for each $i=2, 3, \cdots , s$, $\beta_i\in \beta (\mathcal{C})$ provides an MPP vertex $m_i$, so that Algorithm \ref{alg:MPP-Alg-triangular} must examine all tiles $\beta_{i+1} , \beta_{i+2} ,\cdots , \beta_s$ to find $m_i$. This gives a total number of $1+ (s-2) +(s-3) + \cdots +2+1 = 1+ \frac{1}{2}(s-2)(s-1)$ examinations of tiles, which corresponds to quadratic order. It seems that such a worst case is practically impossible. Figure \ref{fig:complexity} shows an example of backtracking by many tiles, but not immediately repeated for the next MPP vertex to be found. 
Further studies are needed to investigate whether there are complexes where large backtracking on the input list can occur repeatedly for MPP vertices found one after the other.  

\begin{figure}
\centering
\includegraphics[height=2cm]{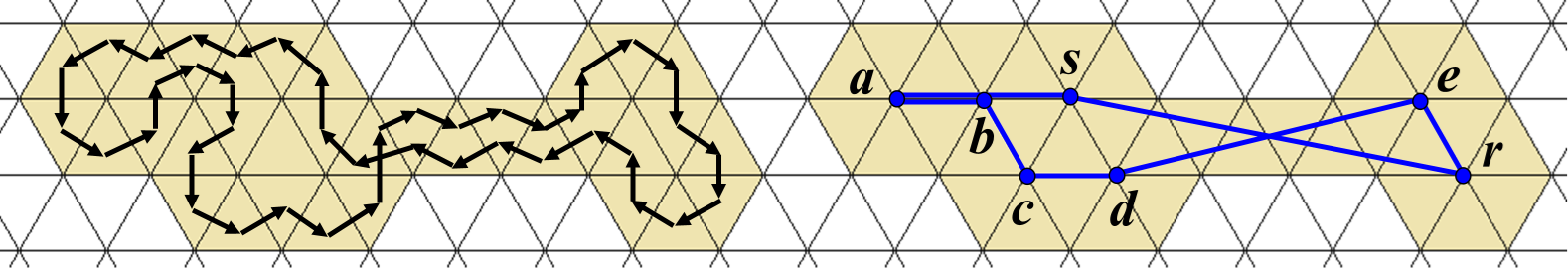}
\caption{With the canonical boundary path of this complex as input, Algorithm \ref{alg:MPP-Alg-triangular} correctly computes the MPP given by the vertex sequence $(a, c, r, e)$. The left figure shows another boundary path which, when used as input list to Algorithm \ref{alg:MPP-Alg-triangular}, produces the sequence of polygonal curve vertices $(a, b, c, d, e, r, s)$. This curve does not correspond to the MPP frontier, which is the shortest polygonal curve that lies inside the complex but circumscribes all core vertices $a, c, d, s, r, e$. The resulting curve does not even describe the frontier of a weakly simple polygon.}
\label{fig:wrong-non-canonical}
\end{figure}

The supposition for Algorithm \ref{alg:MPP-Alg-triangular} that the input list represents the canonical boundary path is essentially important. When using other boundary paths as input, Algorithm \ref{alg:MPP-Alg-triangular} does not necessarily compute the correct MPP vertices, as shown in Figure \ref{fig:wrong-non-canonical}.


\section{Conclusions}
\label{sect:Conclusion}


This article proposes an algorithm to determine the ordered list of vertices of the minimal perimeter polygon (MPP) of any regular complex in the triangular tiling and proves its correctness. The algorithm exploits the special nature of these complexes, which are a type of edge-adjacency connected sets of triangular tiles with no end tiles. The algorithm uses as input data the canonical boun\-dary path, which is easily obtained from the complex via a boundary tracing method also proposed in this work. We show that our MPP algorithm may not correctly work for boundaries that are distinct from the canonical boundary path.

While the boundary tracing algorithm (Algorithm \ref{alg:bound-tracing}) has linear time com\-plexi\-ty depending on the number of boundary tiles, linear com\-plexi\-ty is not achieved for Algorithm \ref{alg:MPP-Alg-triangular}. They may be worst cases where our MPP algorithm requires to examine the boundary tiles a number of times that has quadratic order. Further investigation is needed to optimize the MPP algorithm. 

Under our suppositions, the MPP coincides with the geodesic convex hull, also known as relative convex hull, of a set $A$ with respect to a polygon $B \supset A$, where $A$ can be disconnected. This motivates future work to apply techniques from computational geometry. However, $A$ and $B$ are not explicitly given as input data, these sets must first be determined in an efficient way from the given complex. Other future research may include to analyse the relation between the MPP of digital objects in the triangular tiling and something equivalent to the maximum length digital straight line segments (DSS) known for objects of square pixels, as it is studied in \cite{Lachaud2011LinAlgorMLP,Roussillon2011}. For objects made up of triangular pixels, no DSS has yet been defined in the literature, but first ideas on digital lines were published in \cite{Nagy2017_DSS}.


\section*{References}

\bibliographystyle{splncs04}
\bibliography{Biblio_MPPTriangTiles_July2024}


\end{document}